\documentclass[aps,reprint,english,superscriptaddress,pra]{revtex4-1}
\usepackage[T1]{fontenc}\usepackage{ae} \usepackage{aecompl}
\usepackage{algpseudocode}
\usepackage{tikz}
\usetikzlibrary{quantikz}
\usepackage{rotating}
\usepackage{booktabs}
\usepackage{xcolor}
\usepackage{adjustbox}
\usepackage[utf8]{inputenc}

\usepackage{bbm}
\usepackage{amsmath}
\usepackage{bm}
\usepackage{graphicx}
\usepackage{amssymb}
\usepackage{verbatim}
\usepackage{subfigure}
\usepackage{mathtools}
\usepackage{amssymb}
\usepackage{textcomp}
\usepackage{mathrsfs}
\usepackage[normalem]{ulem}
\usepackage{indentfirst}\usepackage{mathrsfs}
\usepackage{float}
\usepackage{indentfirst}
\usepackage{txfonts}

\emergencystretch = \maxdimen
\makeatletter
\def\journal #1, #2, #3, 1#4#5#6{{\sl #1~}{\bf #2}, #3 (1#4#5#6) }

\makeatother

\usepackage{babel}

\begin{document}
	\title{Probabilistic Nonunitary Gate in Imaginary Time Evolution}
	\author{Tong Liu}
	
	\affiliation{Institute of Physics, Chinese Academy of Sciences, Beijing 100190, China}
	
	\affiliation{School of Physical Sciences, University of Chinese Academy of Sciences, Beijing 100190, China}
	
	\author{Jin-Guo Liu}
	
	\affiliation{Institute of Physics, Chinese Academy of Sciences, Beijing 100190, China}
	
	\author{Heng Fan}
	
	\affiliation{Institute of Physics, Chinese Academy of Sciences, Beijing 100190, China}
	
	\affiliation{School of Physical Sciences, University of Chinese Academy of Sciences, Beijing 100190, China}
	
	\affiliation{CAS Center for Excellence in Topological Quantum Computation, University of Chinese Academy of Sciences, Beijing 100190, China}
	
	\affiliation{Songshan Lake Materials Laboratory, Dongguan 523808, Guangdong, China}

	\begin{abstract}
		
		Simulation of quantum matters is a significant application of quantum computers. 
		In contrast to the unitary operation which can be realized naturally on a quantum computer, the implementation of nonunitary operation, widely used in classical approaches, needs special designing. Here, by application of Grover's algorithm, we extend the probabilistic method of implementing nonunitary operation and show that it can promote success probability without fidelity decreasing. This method can be applied to problems of imaginary time evolution and contraction of tensor networks on a quantum computer.
		
	\end{abstract}
	\maketitle
	
	\section{Introduction}
	
	Tensor network algorithms have been widely used in studying quantum many-body problems. The great success of these methods include, density matrix renormalization group based on matrix product state (MPS)~\cite{Orus2019,Oestlund1995,Vidal2003,Schollwoeck2011} for simulating one-dimensional gapped systems, projected entangled pair states (PEPS) for two-dimensional gapped systems and multi-scale entanglement renormalization ansatz  (MERA) for critical systems~\cite{Cincio2008,Vidal2007,Vidal2008,Verstraete2010,Orus2013}. Additionally, counting problems such as 3-SAT problems and planar graph 3-colorings problems can be converted into tensor network contraction problems~\cite{Biamonte2017,penrose1971applications,Biamonte2014,Johnson2012}. 	However, the realization of some of those methods are still hard for classical computers. The computational complexity of classically simulating PEPS has proven to be \#P-complete~\cite{Schuch2007,PhysRevResearch.2.013010,PhysRevLett.96.220601} and counting perfect matchings in a bipartite graph is also known to be \#P-complete~\cite{valiant1979complexity,Biamonte2017}. 
	
	Nevertheless, the success of some celebrated quantum algorithms, such as Shor algorithm for integer factoring and discrete logarithm~\cite{Shora,Shor1999} and Grover algorithm for searching an item in an unstructured database~\cite{Grover1996,Grover1997}, shows that quantum computers outperform classical computers for some certain problems~\cite{Farhi2001,Farhi2014,Brooke1999,Santoro2002,Grimsley2019}.
    Hence, it is natural to ask whether tensor network algorithms can be simulated on a quantum computer. Though MPS and well-conditioned injective or G-injective PEPS~\cite{SCHUCH20102153} including the resonating valence bond (RVB) state can be prepared on a quantum computer within polynomial time ~\cite{PhysRevLett.95.110503,cramer2010efficient,PhysRevX.5.041044,PhysRevResearch.1.023025,2019arXiv190807958R,PhysRevA.88.032321,PhysRevLett.108.110502}, there still exist some classes of tensor networks belong to BQP-hard~\cite{Arad2008}. 
    
    Recently, some quantum algorithms inspired from nonunitary operation have achieved remarkable progresses. Nonunitary Jastrow operator effectively removes unwanted state and reduces circuit depth in a hybrid quantum-classical approach~\cite{Mazzola2019}. Symmetry-adapted ansatz state shows significant improvement in both fidelity and energy of ground state in variational quantum eigensolver with symmetry projection operator~\cite{Seki2019}. Imaginary time evolution operator cannot be directly simulated on a quantum computer but  can be approximated by unitary operators after Trotter decomposition according to Uhlmann's theorem~\cite{Motta2019,uhlmann1976transition}. The space and time requirements are reduced exponentially compared to classical methods~\cite{Motta2019}. {The eigenvalues of nonunitary matrices can be evaluated by combining phase estimation algorithm with measurement~\cite{Wang2010}}. Nonunitary operators such as creation and annihilation operators have been realized on a quantum simulator~\cite{Kong2019}.
	

	There are some precursive trials of implementing nonunitary operation on quantum computers including probabilistic methods using ancillary qubit~\cite{Williams2004,TERASHIMA2005} and deterministic methods based on the quantum optic system with dissipative elements~\cite{Barnett1998,klyshko1989nonunitary,Roger2015,Tischler2018}.  It has also been proved that an arbitrary nonunitary operation can be converted into a unitary quantum circuit with projection operator~\cite{TERASHIMA2005}.
	Nonunitary operation can also be constructed as a linear combination of unitary operations~\cite{Childs2012,Berry2015}. A promising scheme is embedding an arbitrary nonunitary operator into a larger unitary operator with ancillary qubits in some accuracy $\epsilon$~\cite{Williams2004}. Repeatedly measuring ancillary qubit until it flips marks a successful achievement of desired nonunitary operator. This scheme has an advantage that the failed measurement perturbs the target state little. 
	 
	 Imaginary time evoulution is a powerful tool to calculate the ground state of many-body system ~\cite{Motta2019,McArdle2019} and can be simulated by time-evolving block decimation algorithm (TEBD) on classical computers~\cite{Vidal2003,Vidal2004a}. As a concrete example of tensor network algorithms, we combine the scheme~\cite{Williams2004} with imaginary time evolution algorithm to find the ground state of an Ising Hamiltonian. The result shows that the success probability exponentially decays to zero with lattice size $N$. We use Grover algorithm to boost the success probability of flipping ancillary qubit. The ground state can be prepared on a quantum computer and some interesting quantities such as correlation functions or expectation values can be measured on a real quantum computer.  
	
     In Sec.~\ref{sec:method}, we reivew the method in~\cite{Williams2004} and calculate that the trade-off between fidelity and probability. In Sec~\ref{sec:ite}, we introduce the quantum circuit of imaginary time evolution algorithm and analyze the computational complexity. In Sec.~\ref{sec:grover}, we boost the success probability without lowering the fidelity using Grover's algorithm. All numerical results of this paper is derived from quantum simulator Yao.jl~\cite{YaoFramework2019}.

    \section{Method}\label{sec:method}
	{Given a nonunitary gate $G$, its matrix representation $M$
    $M_{n\times m}$ ($n>m$) can be padded into a square matrix $S_{n\times n}$ and decomposed into two unitary matrices and a diagonal matrix $\Sigma$ by singular value decomposition (SVD).} 
	To simplify our discussion, we begin with diagonal matrix $\Sigma$ are four by four of which diagonal elements $\sigma_i$ are all non-negative. We normalize $\Sigma$ as $\Sigma/||\mathcal{N}||$ such that the norm of each element is less than one. $\mathcal{N}$ is the element with maximum norm in $\Sigma$.  
	For a given operator $\Sigma$ and an initial pure state $|\psi_0\rangle\in\mathcal{H}^2$, our target state is
	\begin{equation}\label{target}
		|\psi\rangle = \frac{\Sigma|\psi_0\rangle}{\sqrt{\langle\psi_0|\Sigma^2|\psi_0\rangle}}.
	\end{equation}
	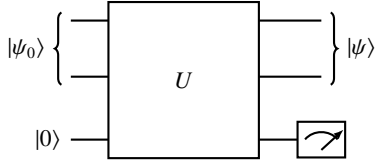
\begin{figure}[h]
		\centering
		\begin{quantikz}
			\lstick[wires = 2]{$\ket{\psi_0}$} & \gate[wires = 3][2cm]{U} & \qw \rstick[wires = 2]{$\ket{\psi}$}\\ & & \qw \\
			\lstick{$\ket{0}$} & & \meter{} 
		\end{quantikz}
	\caption{Three-qubit gate $U$ with two physical qubits and one ancillary qubit.}
	\end{figure}
	 Then we define a three-qubit gate $U(\Sigma,\epsilon)$ as below~\cite{Williams2004}
	\begin{equation}
		U(\Sigma,\epsilon)=\begin{pmatrix}
		\cos(\epsilon\Sigma) & -\sin(\epsilon\Sigma) \\
		\sin(\epsilon\Sigma) &\cos(\epsilon\Sigma)
		\end{pmatrix} = \exp(-i\epsilon(\sigma^y\otimes\Sigma)),
	\end{equation}
	where $\epsilon$ is a very small parameter. The input state $|0\rangle|\psi_0\rangle$ is a product state of an ancillary qubit initialized as $|0\rangle$ and initial state $|\psi_0\rangle$. After applying the 3-qubit gate on the input state we have
	\begin{equation}
		U(|0\rangle|\psi_0\rangle)=|0\rangle\cos(\epsilon\Sigma)|\psi_0\rangle+|1\rangle\sin(\epsilon\Sigma)|\psi_0\rangle.
	\end{equation}
	Now we measure the ancillary qubit. The success probability of getting $|1\rangle$ or flipping ancillary qubit is
	\begin{equation}
		p_0 = \langle\psi_0|\sin^2(\epsilon\Sigma)|\psi_0\rangle,
	\end{equation}
	and the output state is
	\begin{equation}\label{output}
		|\psi'\rangle = \frac{\sin(\epsilon\Sigma)|\psi_0\rangle}{\sqrt{\langle\psi_0|\sin^2(\epsilon\Sigma)|\psi_0\rangle}}.
	\end{equation}
	Obviously,
	\begin{equation}
		\lim_{\epsilon\to 0}|\psi'\rangle = |\psi\rangle
	\end{equation}
	
	If outcome is $|0\rangle$ or ancillary qubit isn't flipped, output state is approximately same as the input state if $\epsilon$ is small enough. We have obtained our goal state otherwise we rotate initial state with angle $O(\epsilon^2)$ if ancillary qubit is flipped after one-shot measurement. Hence we can apply the 3-qubit gate on the output state and measure the ancillary qubit once more until the ancillary qubit is flipped. If lower $\epsilon$, our output state $|\psi'\rangle$ is closer to the target state but the success probability also decreases. There is a trade-off between probability and fidelity. In order to demonstrate the trade-off clearer, we define a quantity $p_\eta(\epsilon)$ which is the success probability when fidelity of output state and target state is greater than $(1-\eta)^2$. We notice that fidelity and probability are both determined by measure times $n$. There is a threshold measurement time $n^*$ corresponding to the fidelity $(1-\eta)^2$. Thus $p_\eta(\epsilon)$ can be calculated as
	\begin{equation}
	\label{p}
		p_\eta(\epsilon) = \sum_{n=1}^{n^*}p(n,\epsilon) = 1-\langle\psi_0|\cos^{2n^*}(\epsilon\Sigma)|\psi_0\rangle,
	\end{equation}
	where $p(n,\epsilon)$ is the success probability after $n$ shots measurement. We give an approximation about $n^*$ in Appendix~\ref{ref:A} where we have divided diagonal matrices into the idempotent matrices $\Sigma^2 = \Sigma$ and the other matrices. The fidelity of idempotent matrices is independent with measure times and is always unity such as a projection operator. We will discuss idempotent matrices in the next section. The threshold measurement times $n^*$ is

	\begin{equation}\label{n}
	n^* \approx \left\lfloor\sqrt{\frac{8\eta \langle\Sigma^2\rangle^2}{\langle\Sigma^6\rangle\langle\Sigma^2\rangle-\langle\Sigma^4\rangle^2}}\frac{1}{\epsilon^2}\right\rfloor .
	\end{equation}
	\begin{figure}
		\centering
		\includegraphics[scale=0.4]{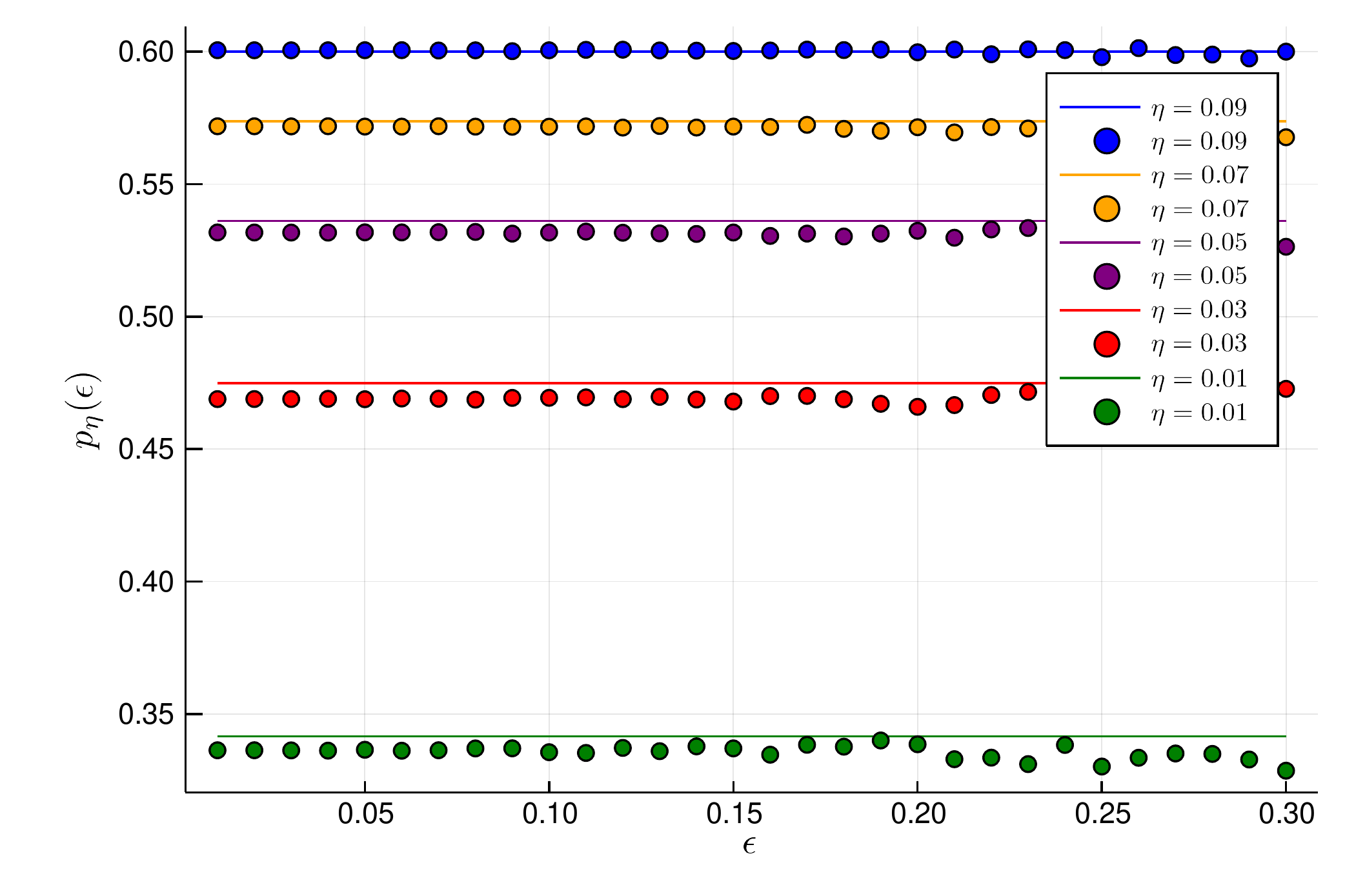}
		\caption{$p_\eta(\epsilon)$ approaches to a constant when $\epsilon$ decreases to zero. Circles are $p_\eta(\epsilon)$ which are sampled from Yao.jl~\cite{YaoFramework2019} and lines are calculated by Eq.~(\ref{peta}). $\Sigma$ is a four by four diagonal matrix of which diagonal element are sampled from $[0,1]$ randomly and input state is the uniformly superposition state.}\label{diffeta}
	\end{figure}

	Now we substitute Eq.~(\ref{n}) into Eq.~(\ref{p}) to calculate success probability using $\lim_{x\to0}\cos^{b/x^2}(ax)=\exp(-a^2b/2)$ and obtain
	\begin{equation}\label{peta}
	p_\eta = \lim_{\epsilon\to 0}p_\eta(\epsilon) =  1 - \left\langle\exp\left(-\sqrt{8\eta}f(\Sigma)\Sigma^2\right)\right\rangle,
	\end{equation}
	where
	\begin{equation}
	f(\Sigma) = \sqrt{\frac{\langle\Sigma^2\rangle^2}{\langle\Sigma^6\rangle\langle\Sigma^2\rangle-\langle\Sigma^4\rangle^2}}.
	\end{equation}
	As shown in Fig.~\ref{diffeta}, $p_\eta(\epsilon)$ will approach to a constant $p_\eta$ when $\epsilon$ approaches to zero. In contrary, $p_\eta = O(\sqrt{\eta})$ when improve fidelity by decreasing $\eta$. It is aslo illustrated by Fig.~\ref{diffeta} where the bottom lines are sparser than top lines. Thus the average measurement times is $O(1/\sqrt{\eta})$. The approximation of $n^*$ and $p_\eta$ can be improved by expanding Taylor series of Eq.(\ref{fidelity}) to higher orders.

	\section{Application:  imaginary time evolution}
	\label{sec:ite}
	\subsection{Quantum circuit of imaginary time evolution}
	Finding a maximum cut of a graph has been proven {NP}-complete~\cite{Garey1974,Papadimitriou1991}. There exists a map from a graph to a classical Ising Hamiltonian which converts a combinatorial optimization problem into a solving the ground state of Ising Hamiltonian problem~\cite{Barahona1988,kirkpatrick1983optimization}. We calculate the ground state of Hamiltonian which can be mapped into a NP-complete problem~\cite{Farhi2014,Garey1974,Papadimitriou1991}
	\begin{equation}
	H = \sum_i J_i\sigma^z_i\sigma^z_{i+1},
	\end{equation}
	where interaction $J_i$ between two qubits is randomly chosen as $1$ or $-1$, and use imaginary time evolution to find the ground state based on nonunitary gates. The imaginary time evolution operator (omitted constant terms) is
	\begin{equation}\label{imag}
	\tilde{U}(\tau) = e^{-H\tau}= \prod_i e^{J_i\sigma_i^z\sigma_{i+1}^z\tau},
	\end{equation}
	where $\tau$ is imaginary time. 
	\begin{figure}[ht]
		\centering
		\begin{quantikz}
			\lstick{$\ket{0}$} & \gate{H} & \gate[wires = 2]{U_{12}(\tau)} & \qw & \qw &\qw \\
			\lstick{$\ket{0}$} & \gate{H} & & \gate[wires = 2]{U_{23}(\tau)} & \qw &\qw\\
			\lstick{$\ket{0}$} & \gate{H} &\qw & & \gate[wires = 2]{U_{34}(\tau)} &\qw\\
			\lstick{$\ket{0}$} & \gate{H} &\qw &\qw & &  \rstick{}\qw
		\end{quantikz}
		\caption{Imaginary time evolution circuit of Ising Hamiltonian with $N = 4$. $|+\rangle = H|0\rangle = (|0\rangle + |1\rangle)/\sqrt{2}$. $H$ is Hadamard gate. }
	\end{figure}
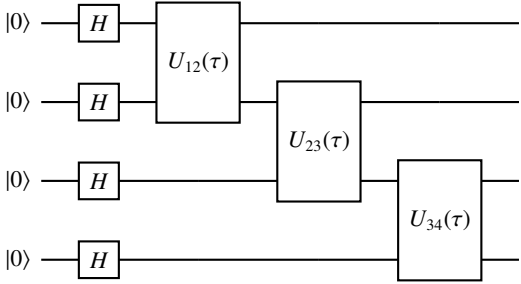

	We can regard each $e^{J_i\sigma_i^z\sigma_{i+1}^z\tau}$ term as a nonunitary operation and realized by a three-qubit gate. Assume each ancillary qubit in three-qubit gates is separable from other ancillary qubit. The product of all unitary gates is
	\begin{align}
	U&=\prod_{ i}\exp\left(-i\epsilon(\sigma^y_{i}\otimes e^{\sigma_i^z\sigma_{i+1}^z\tau})\right) \notag \\
	&=\exp\left(-i\sum_{i}\epsilon(\cosh(\tau)\sigma^y_{i}+\sinh(\tau) \sigma^y_{i}\sigma_i^z\sigma_j^z)\right) \notag \\
	&=\exp\left(-i\epsilon H'\right).\end{align}
	$\sigma_{i}^y$ is applied on the $i$-th ancillary qubit and
	\begin{equation}
	H' = \sum_{i}\sinh(\tau)\sigma_{i}^y\sigma_i^z\sigma_j^z+\cosh(\tau)\sigma^y_{i}.
	\end{equation}
	If cluster-Ising like Hamiltonian $H'$ can be prepared on a quantum computer~\cite{Peng2014,Tseng1999}, $U$ is a real time evolution operator governed by $H'$. After time $\epsilon$, we store a bit string like ``00010010$\dots$" after measuring all ancillary qubits without resetting them. Repeat the procedure including $\epsilon$ time evolution and measurement until all ancillary qubits have been flipped from $|0\rangle$ to $|1\rangle$ at least once. It means all nonunitary operation has been approximately applied at least once. Identity or nonunitary operation are dependent on whether the corresponding ancillary qubit is flipped. 
	
	\subsection{Computational complexity}
	Assume the lattice size is $N+1$ and there are $N$ ancillary qubits and $N$ nonunitary matrices $\Sigma_i$. We denote $n_i$ as the $i$th measurement times to flip the $i$th ancillary qubit. Thus the corresponding $p_\eta(\epsilon)$ is
	\begin{align}\label{npeta}
		p_\eta(\epsilon) &= \sum_{n_1,n_2,\dots,n_N}^{n^*}\langle\psi_0|\prod_{i}^{N}\cos^{2(n_i-1)}(\epsilon\Sigma_i)\sin^2(\epsilon\Sigma_i)|\psi_0\rangle \notag \\
		&=\langle\psi_0|\prod_{i}^{N}(1-\cos^{2n^*}(\epsilon\Sigma_i))|\psi_0\rangle.
	\end{align}
	The threshold measurement times $n^*$ corresponding to the fidelity $(1-\eta)^2$ is
	\begin{equation}
	\label{nstar}
		n^* = O\left(\frac{\sqrt{\eta}}{\epsilon^2}\right).
	\end{equation}
	Substituting Eq.~(\ref{nstar}) into Eq.~(\ref{npeta}) and using $\lim_{x\to 0}\cos^{b/x^2}(ax) = \exp(-a^2b/2)$ again, we obtain that
	\begin{equation}
		p_\eta = O(\eta^{N/2})
	\end{equation}
	The success probability of keeping fidelity greater than $\eta$ decays exponentially with $N$ for a given fidelity. Thus we cannot implement the scheme to prepare ground state on a quantum computer within a polynomial time.

	\section{Combined with Grover's Algorithm}
	\label{sec:grover}
	
\begin{figure}
		\centering
		\subfigure[]{
			\includegraphics[scale=0.30]{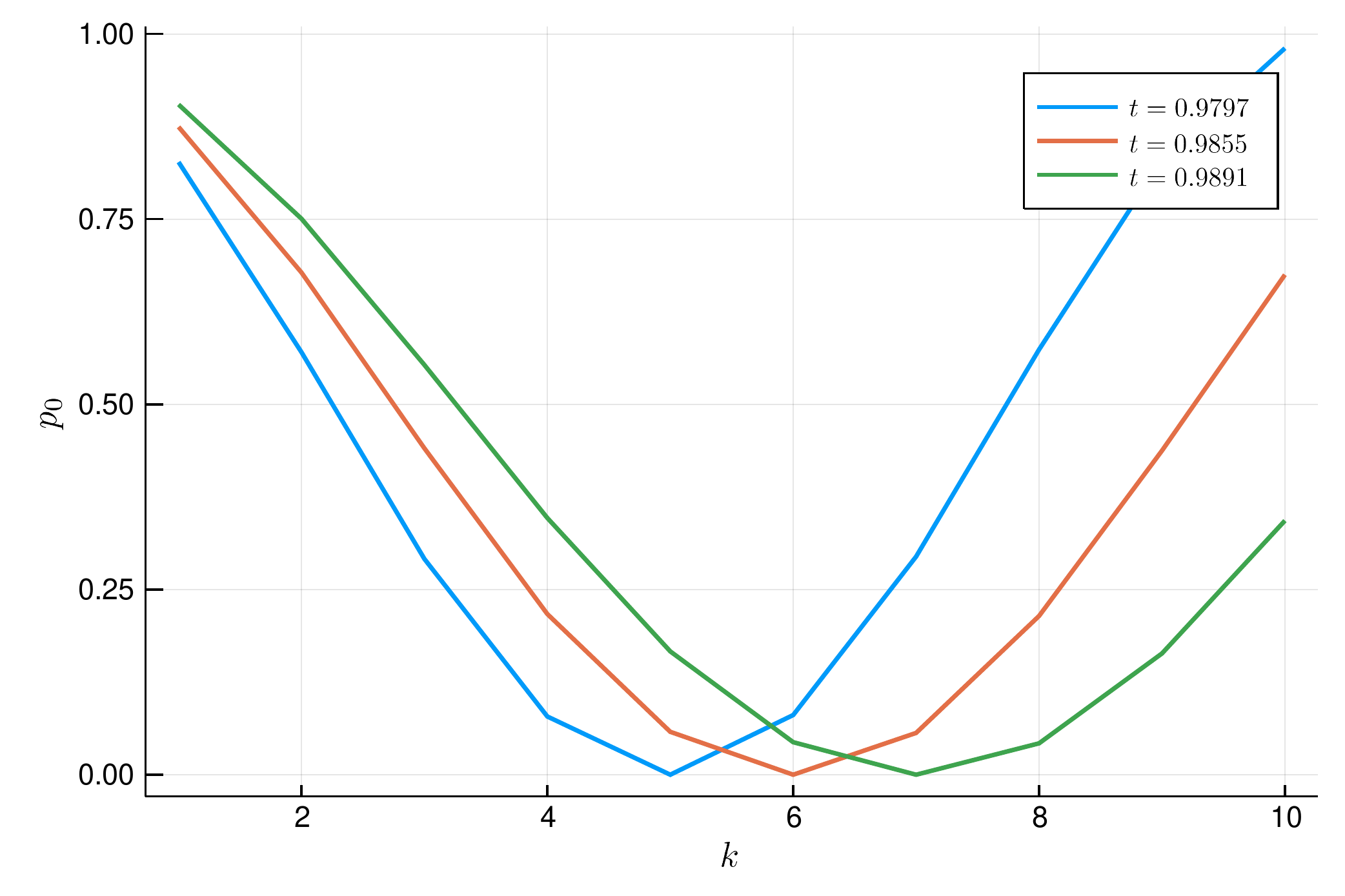}
		}
	\caption{The probability of failure $p_0(t,k)$ verifies the relation between $t^*$ and $k$ in table.}\label{dit}
	\end{figure}

	Grover's algorithm is a quantum search algorithm~\cite{Grover1996,Nielsen:2011:QCQ:1972505} to find solutions for a given question. Many improved quantum search algorithms have been devised~\cite{Cerf1998,Grover2005,Yoder2014,Berry2014}. Fixed-point quantum search algorithm~\cite{Grover2005,Yoder2014} converges to the target state without souffle problem of Grover's origin algorithm. Oblivious amplitude amplification enables us to find target state without reflection operator about initial state. However, the number of Grover iteration steps can be analytically calculated in our problem as we will show. Target state can not be written as $\sqrt{p}U|\phi\rangle$ in~\cite{Berry2014}. Therefore,  we can only apply Grover's origin algorithm in our problem. It consists of Grover iteration $G$ supplied with a quantum oracle to recognize solutions to the search problem.
	Grover iteration $G$ is defined as $G = RO = (2 U|0\rangle\langle 0|_{\rm{anc}} \otimes|\psi_0\rangle\langle\psi_0|U^\dagger-I)O$ where
	\begin{equation}
		O = (|0\rangle\langle 0|-|1\rangle\langle 1|)_{\rm{anc}}\otimes I = Z_{\rm{anc}}\otimes I
	\end{equation}
	The oracle operation $O$ performs a reflection about the $|1\rangle\langle 1|$. Then $ R = 2 U|0\rangle\langle0|_{\rm{anc}}\otimes|\psi_0\rangle\langle\psi_0|U^\dagger-I$ also performs a reflection about $U|0\rangle_{\rm{anc}}|\psi_0\rangle$ in the plane. The two successive reflections take state $U|0\rangle_{\rm{anc}}|\psi_0\rangle$ to
	\begin{equation}
		\alpha^{(1)}_1|0\rangle_{\rm{anc}}\cos(\epsilon\Sigma)|\psi_0\rangle \notag 
		+ \alpha^{(1)}_2|1\rangle_{\rm{anc}}\sin(\epsilon\Sigma)|\psi_0\rangle
	\end{equation} 
	where
	\begin{gather}
		\alpha^{(1)}_1 = 2\langle\psi_0|\cos(2\epsilon\Sigma)|\psi_0\rangle - 1,  \\ \alpha^{(1)}_2 = 2\langle\psi_0|\cos(2\epsilon\Sigma)|\psi_0\rangle + 1
	\end{gather}
	 It follows that continued application of $G$ takes the state to 
	 \begin{equation}
	 	\alpha^{(k)}_1|0\rangle_{\rm{anc}}\cos(\epsilon\Sigma)|\psi_0\rangle+\alpha^{(k)}_2|1\rangle_{\rm{anc}}\sin(\epsilon\Sigma)|\psi_0\rangle 
	 \end{equation}
	 The relation between $(\alpha^{(k)}_1,\alpha^{(k)}_2)$ and $(\alpha^{(k+1)}_1,\alpha^{(k+1)}_2)$ is
	 \begin{equation}
	 	\begin{pmatrix}
	 	\alpha^{(k+1)}_1 \\
	 	\alpha^{(k+1)}_2
	 	\end{pmatrix}
	 	= T
	 	\begin{pmatrix}
	 	\alpha^{(k)}_1 \\ \alpha^{(k)}_2   
	 	\end{pmatrix}
	 \end{equation}
	where
	\begin{equation}
		T = \begin{pmatrix}
		2\langle\psi_0|\cos^2(\epsilon\Sigma)|\psi_0\rangle-1 & -2\langle\psi_0|\sin^2(\epsilon\Sigma)|\psi_0\rangle \\
		2\langle\psi_0|\cos^2(\epsilon\Sigma)|\psi_0\rangle & -2\langle\psi_0|\sin^2(\epsilon\Sigma)|\psi_0\rangle +1 
		\end{pmatrix}
	\end{equation}
	$T$ is called transfer matrix here. Now the failure probability is
	 \begin{align}\label{p0}
	 	p_0(t,k) & 
	 	= \frac{1}{4}\left[\left(\sqrt{t}+\sqrt{t-1}\right)^{2k+1}+\left(\sqrt{t}-\sqrt{t-1}\right)^{2k+1}\right]^2
	 \end{align}
	 where $t = \langle\psi_0|\cos^2(\epsilon\Sigma)|\psi_0\rangle$. Due to $0<t<1$, $\sqrt{t}+\sqrt{t-1}$ is a complex number $\in U(1)$. Suppose the argument of it is $\theta$, then we have the failure probability
	 \begin{equation}
	 	p_0(\theta,k) = \cos^2\left((2k+1)\theta\right).
	 \end{equation}
	The optimal $t^*$ is the root of $p_0(t,k)$. Then we obtain
	\begin{equation}\label{eq:root}
	 	t^*_m = \cos^2\frac{(2m+1)\pi}{4k+2},\quad m \in \mathbb{Z}.
	\end{equation} 
	 Some roots from Eq.(\ref{eq:root}) are shown in Fig.\ref{tab}. We simulated the algorithm in Yao.jl and plot data when $m=0$ and $t = 0.9797$, $0.9855$ and $0.9891$ in Fig.\ref{dit}. And the optimal times of Grover iteration are 5, 6 and 7 respectively as we have proposed.
	\begin{figure}[h]
		\centering
		\begin{tabular}{c|c|c|c|c|c|c|c}
			\hline
			$k$   & 1 & 2 &3 &4 &5 &6 &7 \\ \hline
			$t^*_0$ & 0.7500 & 0.9045 & 0.9505 & 0.9698 &0.9797 &0.9855 &0.9891 \\
			\hline
		\end{tabular}
		\caption{Roots $t^*$ of Eq.(24) for different number of Grover iterations $k$.}\label{tab}
	\end{figure}

	After $k$ Grover iterations, ancillary qubit will be flipped after one measurement if $\alpha_k = 0$. Now the fidelity is totally determined by $\epsilon$. According to Eq.(\ref{fidelity}), we can see that the fidelity will be 1 independent with $\epsilon$ if $\Sigma$ such as projection operator saturates the inequality. That's to say, it is possible to apply only \emph{one} Grover iteration to take our output state into the target state.
	
	There are two methods to implement imaginary time evolution operator $\tilde{U}$ based on the Grover's algorithm. 
	
	\subsection{Method (\romannumeral1)}
	Suppose the Hamiltonian is written as $H =-\sum_i J_i\sigma_i\sigma_{i+1}$. The first method is that divide $\tilde{U} = e^{-H\tau}$ into the product of $N$ local terms. Each of them can be realized by nonunitary gate and boosted by Grover's algorithm. Here,
	\begin{equation}\label{sigma}
	\Sigma_i(\tau) 
	= \begin{dcases}
	\mathrm{diag}(1,\exp(-2 J_i \tau),\exp(-2 J_i\tau),1), & J_i > 0; \\
	\mathrm{diag}(\exp(-2 J_i \tau),1,1,\exp(-2 J_i \tau)), & J_i <0.
	\end{dcases} 
	\end{equation}
	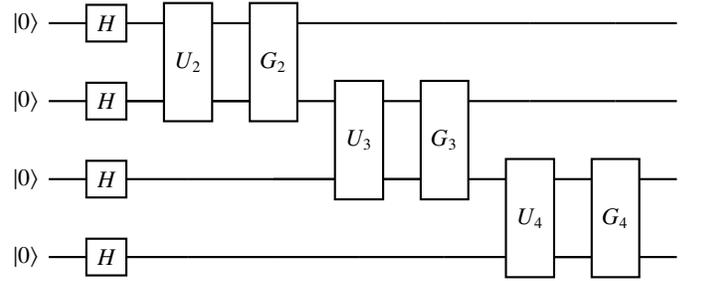
\begin{figure}[h]\label{m1}
		\centering
		\begin{quantikz}
			\lstick{$\ket{0}$} & \gate{H} & \gate[wires = 2]{U_2} & \gate[wires = 2]{G_2} & \qw & \qw & \qw &\qw &\qw \\
			\lstick{$\ket{0}$} & \gate{H} & \qw  &\qw  &\gate[wires = 2]{U_{3}} &\gate[wires=2]{G_3} & \qw & \qw &\qw \\
			\lstick{$\ket{0}$} & \gate{H} & \qw  &\qw & \qw & \qw & \gate[wires = 2]{U_4} &\gate[wires = 2]{G_4} &\qw \\
			\lstick{$\ket{0}$} & \gate{H} & \qw & \qw & \qw & \qw & & \qw &\qw
		\end{quantikz}
	\caption{The first way to implement Grover's algorithm in imaginary time evolution. Ancillary qubits are absorbed into $U$ and $G$.}
	\end{figure}

	One important limitation of Grover's algorithm is that either there are a great many replicas of input state or input state can be prepared clearly. The input states of nonunitary gates except the first one are unknown to us which is the output state of the former one. For example, the input state of $U(\Sigma_i,\epsilon_i)$ is the tensor product of output state of last qubit and a new single qubit in imaginary time evolution algorithm i.e., 
	\begin{equation}
		\frac{\prod_{j=1}^{i-1}\sin(\epsilon_j\Sigma_j)H^{\otimes{i}}|0\rangle}{\sqrt{\langle 0|H^{\otimes i}\prod_{j=1}^{i-1}\sin^2(\epsilon_j\Sigma_j)H^{\otimes{i}}|0\rangle}}\otimes H|0\rangle
	\end{equation}
	
	 Thus it is necessary to demonstrate that how to construct the reflection operator which relies on the input state in each Grover iteration. Denote $|\psi_i\rangle\in \mathcal{H}^{\otimes(i+1)}$ as the output state of the $i$ th $U(\Sigma_i,\epsilon_i)$ where $U(\Sigma_i,\epsilon_i)$ acts on the $(i-1)$ th and $i$ th physical qubit. Hence,
	\begin{equation}
		|\psi_i\rangle = \left[I_{\rm{anc}}\otimes I_{i-1}\otimes U(\Sigma_i,\epsilon_i)\right]|0\rangle_{\rm{anc}}\otimes G_{i-1}|\psi_{i-1}\rangle\otimes H_{i+1}|0\rangle_{i+1}
	\end{equation}
	 where $I_{i-1}$ is the identity operator defining in the Hilbert space of the first $i-1$ qubit and $G_{i-1}$ is the $(i-1)$ th Grover iteration. Reflection operator $R_i = 2|\psi_i\rangle\langle\psi_i|-I$ appeared in the $i$ th Grover iteration and $G_i = R_i O_i = (2|\psi_i\rangle\langle\psi_i|-I)Z_{i+1}$. For simplicity, we denote $U_i = U(\Sigma_i,\epsilon_i)$ and $|\psi_0\rangle = |+\rangle^{\otimes N} = H^{\otimes i}|0\rangle$. Thus $R_i$ can be constructed as shown in Appendix \ref{sec:ri}.

	It is sufficient to apply Grover iteration once where $t = 0.75$ in each local term because $\Sigma_i(\tau)$ approximately saturates $\Sigma_i^2 = \Sigma_i$. The fidelity of each nonunitary gate is $0.9999998$.

	The oracle in each Grover iteration is always $Z$ gate applied in the ancillary qubit. The number of gate in $R_i$ is roughly three times of that in $R_{i-1}$. Thus the number of gates in the $i$-th iteration is $O(2^i)$ and total number of gates consisting of Grover iterations which are used to find ground state of Hamiltonian composed of nearest-neighbour interaction is $O(2^N)$ with $O(N)$ qubits. 

	\subsection{Method (\romannumeral2)}
	\begin{figure}[h]
		\centering
		\begin{quantikz}
			\lstick{$\ket{0}$} & \gate{H} &\gate[wires=2]{U_2} &\qw & \qw & \gate[wires=4]{G} &\qw \\
			\lstick{$\ket{0}$} & \gate{H} &\qw & \gate[wires=2]{U_3} & \qw & & \qw \\
			\lstick{$\ket{0}$} & \gate{H} &\qw & \qw & \gate[wires=2]{U_4} &  & \qw \\
			\lstick{$\ket{0}$} & \gate{H} & \qw & \qw & \qw & & \qw
		\end{quantikz}
		\caption{The second method to implement Grover's algorithm in imaginary time evolution.}
	\end{figure}
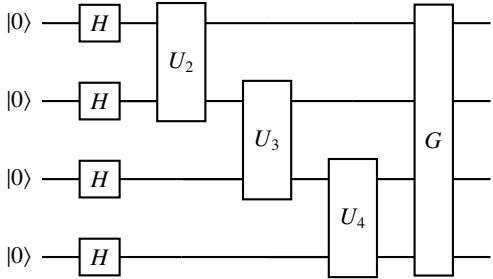
\begin{figure}
		\centering
		\includegraphics[scale=0.4]{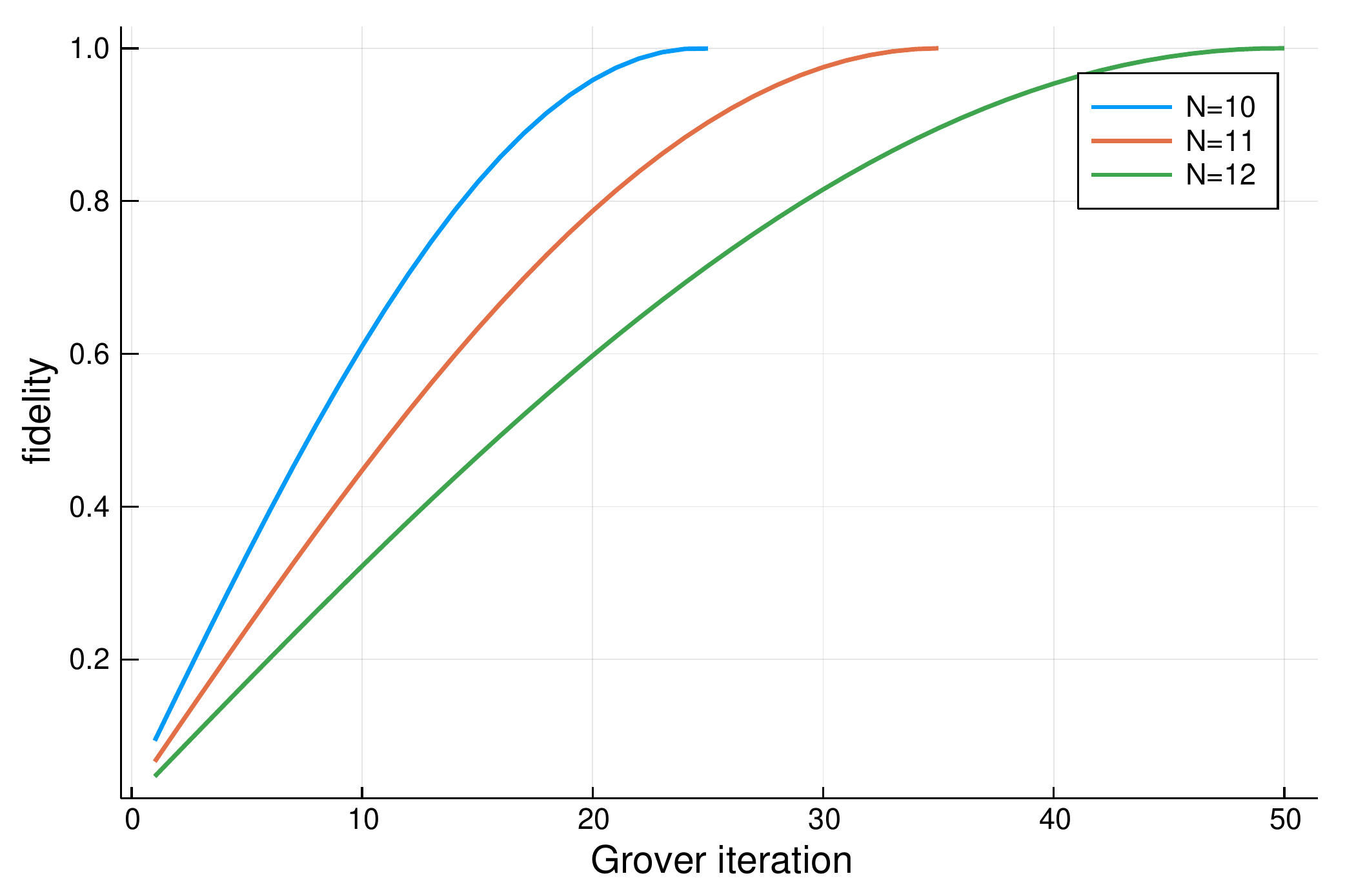}
		\caption{Fidelity between output state and ground state of  after Grover iteration. The lattice size $N=10,11$ and $12$ and imaginary time evolution time $\tau = 10$.}\label{groundstate}
	\end{figure}
\begin{figure}[h]
		\centering
		\includegraphics[scale=0.4]{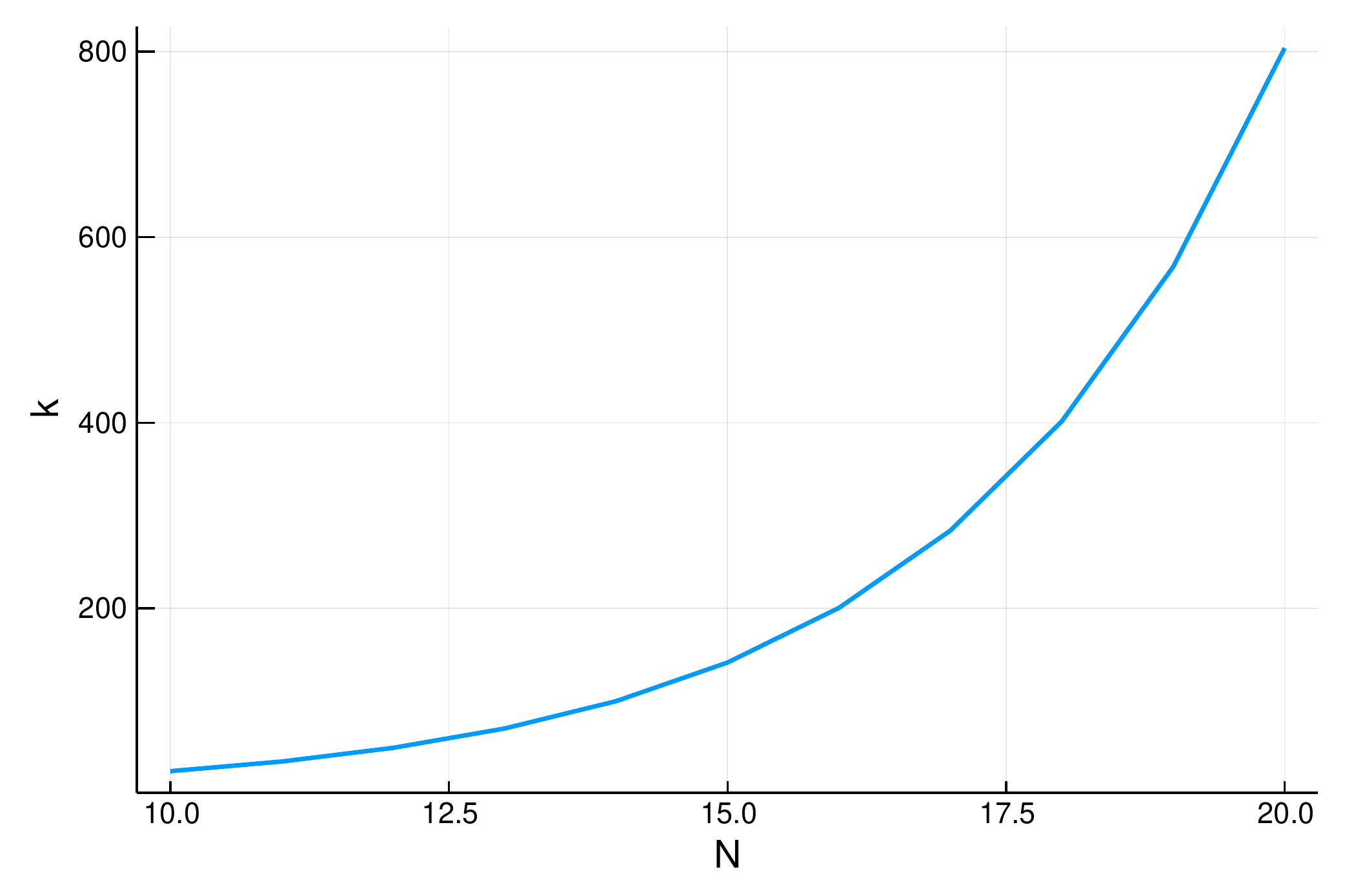}
		\caption{The relation between Grover iteration $k$ and lattice size $N$.}\label{k}
	\end{figure}

	The second method is that implement the whole $\tilde{U}$ with cluster-Ising like Hamiltonian as we have proposed in Sec \ref{sec:ite}. The search space is the Hilbert space $\mathcal{H}^{\otimes{N-1}}$ composed of $(N-1)$ ancillary qubits. We want to apply Grover's algorithm to find the state $|1^{N-1}\rangle$. Now the input state $|\psi_0\rangle $ is $I_{\rm{anc}}\otimes H^{\otimes N-1}|0\rangle_{\rm{anc}}\otimes |0\rangle$. After $\epsilon$ time, the state evolves into
	\begin{equation}
		|\psi\rangle = |1^{N-1}\rangle \prod_{i=1}^{N-1} \sin(\epsilon\Sigma_i)|\psi_0\rangle + |\phi\rangle.
	\end{equation}	
	where $(|1\rangle\langle 1|_\mathrm{anc}\otimes I)|\phi\rangle = 0$. Success probability after $k$ Grover iterations is
	\begin{equation}\label{p1}
		p_1(s,k) = \frac{1}{4}\left[\left(\sqrt{s}+\sqrt{s-1}\right)^{2k+1}+\left(\sqrt{s}-\sqrt{s-1}\right)^{2k+1}\right]^2
	\end{equation}
	where $s$ is denoted by $\langle\psi_0|\prod_{j=1}^{N-1}\sin^2(\epsilon\Sigma_j)|\psi_0\rangle$
	(the proof of Eq.(\ref{p1}) in Appendix \ref{sec:tm}). It seems as similar as the Eq.(\ref{p0}).  Let $p_1(s,k) = 1$ and we obtain
	\begin{equation}\label{sm}
		s^*_m = \cos^2\frac{m\pi}{2k+1},\quad m\in\mathbb{Z}
	\end{equation}
	Combined with $|\psi_0\rangle=H^{\otimes N}|0\rangle$ and $\Sigma_i(\tau) = e^{J_i\sigma_{i+1}\sigma_i\tau}$, $\epsilon$ can be determined by solving
	\begin{equation}\label{rotzz}
		s^* = \frac{1}{2^{N-1}}\prod_j^{N-1}\left[\sin^2(\epsilon\exp(J_i\tau))+\sin^2(\epsilon\exp(-J_i\tau))\right]
	\end{equation}
	Let $m = k$ in Eq.(\ref{sm}). We obtain
	\begin{equation}\label{sk}
		s_k^* = \cos^2\left(\frac{\pi}{2}-\frac{\pi}{4k+2}\right) = \sin^2\frac{\pi}{4k+2}
	\end{equation}
	$k$ can be approximated as $O(2^{N/2})$ by combining Eq.(\ref{sk}) with Eq.(\ref{rotzz}) which shows quadratic acceleration of Grover's algorithm. The number of Grover iteration exponentially increases with the lattice size $N$ as shown in Fig.~\ref{groundstate} and Fig.~\ref{k}. 
	
	\section{conclusions}
		We testify the repeat-until-success scheme of achieving nonunitary operator proposed in~\cite{Williams2004} and find that the scheme cannot prepare high fidelity nonunitary operator within $O(\mathrm{poly}(N))$ time. The accumulated distortion during each unsuccessful measurement cannot be ignored until the ancillary qubit is flipped. 
		
		With the help of Grover algorithm, we improve the success probability by introducing two methods. Both of them can improve the success probability and fidelity to unity. The number of gates in method (\romannumeral1) may be decreased to $O(\mathrm{poly}(N))$ by assuming the entanglement of ground state is short-distance with loss of fidelity. In method (\romannumeral2), our target state lies in a Hilbert space consisting of $N$ ancillary qubits, thus the complexity is $O(2^{N/2})$. We also calculate the ground state of Hamiltonian where appears in the combinatorial optimization problem using nonunitary operation and imaginary time evolution. 
		
	\begin{acknowledgments}
	
	This work was supported by the National Key R \& D Program of China (Grants No. 2016YFA0302104 and No. 2016YFA0300600), the National Natural Science Foundation of China (Grants No. 11774406 and No. 11934018), the Strategic Priority Research Program of the Chinese Academy of Sciences (Grant No. XDB28000000), and the Beijing Academy of Quantum Information Science (Grant No. Y18G07).
	\end{acknowledgments}

	\bibliography{ref}

\begin{thebibliography}{63}%
\makeatletter
\providecommand \@ifxundefined [1]{%
 \@ifx{#1\undefined}
}%
\providecommand \@ifnum [1]{%
 \ifnum #1\expandafter \@firstoftwo
 \else \expandafter \@secondoftwo
 \fi
}%
\providecommand \@ifx [1]{%
 \ifx #1\expandafter \@firstoftwo
 \else \expandafter \@secondoftwo
 \fi
}%
\providecommand \natexlab [1]{#1}%
\providecommand \enquote  [1]{``#1''}%
\providecommand \bibnamefont  [1]{#1}%
\providecommand \bibfnamefont [1]{#1}%
\providecommand \citenamefont [1]{#1}%
\providecommand \href@noop [0]{\@secondoftwo}%
\providecommand \href [0]{\begingroup \@sanitize@url \@href}%
\providecommand \@href[1]{\@@startlink{#1}\@@href}%
\providecommand \@@href[1]{\endgroup#1\@@endlink}%
\providecommand \@sanitize@url [0]{\catcode `\\12\catcode `\$12\catcode
  `\&12\catcode `\#12\catcode `\^12\catcode `\_12\catcode `\%12\relax}%
\providecommand \@@startlink[1]{}%
\providecommand \@@endlink[0]{}%
\providecommand \url  [0]{\begingroup\@sanitize@url \@url }%
\providecommand \@url [1]{\endgroup\@href {#1}{\urlprefix }}%
\providecommand \urlprefix  [0]{URL }%
\providecommand \Eprint [0]{\href }%
\providecommand \doibase [0]{https://doi.org/}%
\providecommand \selectlanguage [0]{\@gobble}%
\providecommand \bibinfo  [0]{\@secondoftwo}%
\providecommand \bibfield  [0]{\@secondoftwo}%
\providecommand \translation [1]{[#1]}%
\providecommand \BibitemOpen [0]{}%
\providecommand \bibitemStop [0]{}%
\providecommand \bibitemNoStop [0]{.\EOS\space}%
\providecommand \EOS [0]{\spacefactor3000\relax}%
\providecommand \BibitemShut  [1]{\csname bibitem#1\endcsname}%
\let\auto@bib@innerbib\@empty
\bibitem [{\citenamefont {Or{\'{u}}s}(2019)}]{Orus2019}%
  \BibitemOpen
  \bibfield  {author} {\bibinfo {author} {\bibfnamefont {R.}~\bibnamefont
  {Or{\'{u}}s}},\ }\bibfield  {title} {\bibinfo {title} {Tensor networks for
  complex quantum systems},\ }\href {https://doi.org/10.1038/s42254-019-0086-7}
  {\bibfield  {journal} {\bibinfo  {journal} {Nature Reviews Physics}\ }\textbf
  {\bibinfo {volume} {1}},\ \bibinfo {pages} {538} (\bibinfo {year}
  {2019})}\BibitemShut {NoStop}%
\bibitem [{\citenamefont {Östlund}\ and\ \citenamefont
  {Rommer}(1995)}]{Oestlund1995}%
  \BibitemOpen
  \bibfield  {author} {\bibinfo {author} {\bibfnamefont {S.}~\bibnamefont
  {Östlund}}\ and\ \bibinfo {author} {\bibfnamefont {S.}~\bibnamefont
  {Rommer}},\ }\bibfield  {title} {\bibinfo {title} {Thermodynamic limit of
  density matrix renormalization},\ }\href
  {https://doi.org/10.1103/PhysRevLett.75.3537} {\bibfield  {journal} {\bibinfo
   {journal} {Physical Review Letters}\ }\textbf {\bibinfo {volume} {75}},\
  \bibinfo {pages} {3537} (\bibinfo {year} {1995})}\BibitemShut {NoStop}%
\bibitem [{\citenamefont {Vidal}(2003)}]{Vidal2003}%
  \BibitemOpen
  \bibfield  {author} {\bibinfo {author} {\bibfnamefont {G.}~\bibnamefont
  {Vidal}},\ }\bibfield  {title} {\bibinfo {title} {Efficient classical
  simulation of slightly entangled quantum computations},\ }\bibfield
  {journal} {\bibinfo  {journal} {Physical Review Letters}\ }\textbf {\bibinfo
  {volume} {91}},\ \href {https://doi.org/10.1103/PhysRevLett.91.147902}
  {10.1103/PhysRevLett.91.147902} (\bibinfo {year} {2003})\BibitemShut
  {NoStop}%
\bibitem [{\citenamefont {Schollwöck}(2011)}]{Schollwoeck2011}%
  \BibitemOpen
  \bibfield  {author} {\bibinfo {author} {\bibfnamefont {U.}~\bibnamefont
  {Schollwöck}},\ }\bibfield  {title} {\bibinfo {title} {The density-matrix
  renormalization group in the age of matrix product states},\ }\href
  {https://doi.org/10.1016/j.aop.2010.09.012} {\bibfield  {journal} {\bibinfo
  {journal} {Annals of Physics}\ }\textbf {\bibinfo {volume} {326}},\ \bibinfo
  {pages} {96} (\bibinfo {year} {2011})}\BibitemShut {NoStop}%
\bibitem [{\citenamefont {Cincio}\ \emph {et~al.}(2008)\citenamefont {Cincio},
  \citenamefont {Dziarmaga},\ and\ \citenamefont {Rams}}]{Cincio2008}%
  \BibitemOpen
  \bibfield  {author} {\bibinfo {author} {\bibfnamefont {L.}~\bibnamefont
  {Cincio}}, \bibinfo {author} {\bibfnamefont {J.}~\bibnamefont {Dziarmaga}},\
  and\ \bibinfo {author} {\bibfnamefont {M.~M.}\ \bibnamefont {Rams}},\
  }\bibfield  {title} {\bibinfo {title} {Multiscale entanglement
  renormalization ansatz in two dimensions: Quantum ising model},\ }\bibfield
  {journal} {\bibinfo  {journal} {Physical Review Letters}\ }\textbf {\bibinfo
  {volume} {100}},\ \href {https://doi.org/10.1103/PhysRevLett.100.240603}
  {10.1103/PhysRevLett.100.240603} (\bibinfo {year} {2008})\BibitemShut
  {NoStop}%
\bibitem [{\citenamefont {Vidal}(2007)}]{Vidal2007}%
  \BibitemOpen
  \bibfield  {author} {\bibinfo {author} {\bibfnamefont {G.}~\bibnamefont
  {Vidal}},\ }\bibfield  {title} {\bibinfo {title} {Entanglement
  renormalization},\ }\bibfield  {journal} {\bibinfo  {journal} {Physical
  Review Letters}\ }\textbf {\bibinfo {volume} {99}},\ \href
  {https://doi.org/10.1103/physrevlett.99.220405}
  {10.1103/physrevlett.99.220405} (\bibinfo {year} {2007})\BibitemShut
  {NoStop}%
\bibitem [{\citenamefont {Vidal}(2008)}]{Vidal2008}%
  \BibitemOpen
  \bibfield  {author} {\bibinfo {author} {\bibfnamefont {G.}~\bibnamefont
  {Vidal}},\ }\bibfield  {title} {\bibinfo {title} {Class of quantum many-body
  states that can be efficiently simulated},\ }\bibfield  {journal} {\bibinfo
  {journal} {Physical Review Letters}\ }\textbf {\bibinfo {volume} {101}},\
  \href {https://doi.org/10.1103/PhysRevLett.101.110501}
  {10.1103/PhysRevLett.101.110501} (\bibinfo {year} {2008})\BibitemShut
  {NoStop}%
\bibitem [{\citenamefont {Verstraete}\ and\ \citenamefont
  {Cirac}(2010)}]{Verstraete2010}%
  \BibitemOpen
  \bibfield  {author} {\bibinfo {author} {\bibfnamefont {F.}~\bibnamefont
  {Verstraete}}\ and\ \bibinfo {author} {\bibfnamefont {J.~I.}\ \bibnamefont
  {Cirac}},\ }\bibfield  {title} {\bibinfo {title} {Continuous matrix product
  states for quantum fields},\ }\bibfield  {journal} {\bibinfo  {journal}
  {Physical Review Letters}\ }\textbf {\bibinfo {volume} {104}},\ \href
  {https://doi.org/10.1103/PhysRevLett.104.190405}
  {10.1103/PhysRevLett.104.190405} (\bibinfo {year} {2010})\BibitemShut
  {NoStop}%
\bibitem [{\citenamefont {Orus}()}]{Orus2013}%
  \BibitemOpen
  \bibfield  {author} {\bibinfo {author} {\bibfnamefont {R.}~\bibnamefont
  {Orus}},\ }\bibfield  {title} {\bibinfo {title} {A practical introduction to
  tensor networks: Matrix product states and projected entangled pair states}\
  }\href {https://doi.org/10.1016/j.aop.2014.06.013}
  {10.1016/j.aop.2014.06.013},\ \Eprint
  {https://arxiv.org/abs/http://arxiv.org/abs/1306.2164v3}
  {http://arxiv.org/abs/1306.2164v3} \BibitemShut {NoStop}%
\bibitem [{\citenamefont {Biamonte}\ and\ \citenamefont
  {Bergholm}()}]{Biamonte2017}%
  \BibitemOpen
  \bibfield  {author} {\bibinfo {author} {\bibfnamefont {J.}~\bibnamefont
  {Biamonte}}\ and\ \bibinfo {author} {\bibfnamefont {V.}~\bibnamefont
  {Bergholm}},\ }\bibfield  {title} {\bibinfo {title} {Tensor networks in a
  nutshell},\ }\href {http://arxiv.org/abs/1708.00006v1} {\ }\Eprint
  {https://arxiv.org/abs/arXiv:1708.0006v1} {arXiv:1708.0006v1} \BibitemShut
  {NoStop}%
\bibitem [{\citenamefont {Penrose}(1971)}]{penrose1971applications}%
  \BibitemOpen
  \bibfield  {author} {\bibinfo {author} {\bibfnamefont {R.}~\bibnamefont
  {Penrose}},\ }\bibfield  {title} {\bibinfo {title} {Applications of negative
  dimensional tensors},\ }\href@noop {} {\bibfield  {journal} {\bibinfo
  {journal} {Combinatorial mathematics and its applications}\ }\textbf
  {\bibinfo {volume} {1}},\ \bibinfo {pages} {221} (\bibinfo {year}
  {1971})}\BibitemShut {NoStop}%
\bibitem [{\citenamefont {Biamonte}\ \emph {et~al.}()\citenamefont {Biamonte},
  \citenamefont {Morton},\ and\ \citenamefont {Turner}}]{Biamonte2014}%
  \BibitemOpen
  \bibfield  {author} {\bibinfo {author} {\bibfnamefont {J.~D.}\ \bibnamefont
  {Biamonte}}, \bibinfo {author} {\bibfnamefont {J.}~\bibnamefont {Morton}},\
  and\ \bibinfo {author} {\bibfnamefont {J.~W.}\ \bibnamefont {Turner}},\
  }\bibfield  {title} {\bibinfo {title} {Tensor network contractions for
  \#sat}\ }\href {https://doi.org/10.1007/s10955-015-1276-z}
  {10.1007/s10955-015-1276-z},\ \Eprint
  {https://arxiv.org/abs/http://arxiv.org/abs/1405.7375v2}
  {http://arxiv.org/abs/1405.7375v2} \BibitemShut {NoStop}%
\bibitem [{\citenamefont {Johnson}\ \emph {et~al.}()\citenamefont {Johnson},
  \citenamefont {Biamonte}, \citenamefont {Clark},\ and\ \citenamefont
  {Jaksch}}]{Johnson2012}%
  \BibitemOpen
  \bibfield  {author} {\bibinfo {author} {\bibfnamefont {T.~H.}\ \bibnamefont
  {Johnson}}, \bibinfo {author} {\bibfnamefont {J.~D.}\ \bibnamefont
  {Biamonte}}, \bibinfo {author} {\bibfnamefont {S.~R.}\ \bibnamefont
  {Clark}},\ and\ \bibinfo {author} {\bibfnamefont {D.}~\bibnamefont
  {Jaksch}},\ }\bibfield  {title} {\bibinfo {title} {Solving search problems by
  strongly simulating quantum circuits}\ }\href
  {https://doi.org/10.1038/srep01235} {10.1038/srep01235},\ \Eprint
  {https://arxiv.org/abs/http://arxiv.org/abs/1209.6010v1}
  {http://arxiv.org/abs/1209.6010v1} \BibitemShut {NoStop}%
\bibitem [{\citenamefont {Schuch}\ \emph {et~al.}(2007)\citenamefont {Schuch},
  \citenamefont {Wolf}, \citenamefont {Verstraete},\ and\ \citenamefont
  {Cirac}}]{Schuch2007}%
  \BibitemOpen
  \bibfield  {author} {\bibinfo {author} {\bibfnamefont {N.}~\bibnamefont
  {Schuch}}, \bibinfo {author} {\bibfnamefont {M.~M.}\ \bibnamefont {Wolf}},
  \bibinfo {author} {\bibfnamefont {F.}~\bibnamefont {Verstraete}},\ and\
  \bibinfo {author} {\bibfnamefont {J.~I.}\ \bibnamefont {Cirac}},\ }\bibfield
  {title} {\bibinfo {title} {Computational complexity of projected entangled
  pair states},\ }\bibfield  {journal} {\bibinfo  {journal} {Physical Review
  Letters}\ }\textbf {\bibinfo {volume} {98}},\ \href
  {https://doi.org/10.1103/PhysRevLett.98.140506}
  {10.1103/PhysRevLett.98.140506} (\bibinfo {year} {2007})\BibitemShut
  {NoStop}%
\bibitem [{\citenamefont {Haferkamp}\ \emph {et~al.}(2020)\citenamefont
  {Haferkamp}, \citenamefont {Hangleiter}, \citenamefont {Eisert},\ and\
  \citenamefont {Gluza}}]{PhysRevResearch.2.013010}%
  \BibitemOpen
  \bibfield  {author} {\bibinfo {author} {\bibfnamefont {J.}~\bibnamefont
  {Haferkamp}}, \bibinfo {author} {\bibfnamefont {D.}~\bibnamefont
  {Hangleiter}}, \bibinfo {author} {\bibfnamefont {J.}~\bibnamefont {Eisert}},\
  and\ \bibinfo {author} {\bibfnamefont {M.}~\bibnamefont {Gluza}},\ }\bibfield
   {title} {\bibinfo {title} {Contracting projected entangled pair states is
  average-case hard},\ }\href
  {https://doi.org/10.1103/PhysRevResearch.2.013010} {\bibfield  {journal}
  {\bibinfo  {journal} {Phys. Rev. Research}\ }\textbf {\bibinfo {volume}
  {2}},\ \bibinfo {pages} {013010} (\bibinfo {year} {2020})}\BibitemShut
  {NoStop}%
\bibitem [{\citenamefont {Verstraete}\ \emph {et~al.}(2006)\citenamefont
  {Verstraete}, \citenamefont {Wolf}, \citenamefont {Perez-Garcia},\ and\
  \citenamefont {Cirac}}]{PhysRevLett.96.220601}%
  \BibitemOpen
  \bibfield  {author} {\bibinfo {author} {\bibfnamefont {F.}~\bibnamefont
  {Verstraete}}, \bibinfo {author} {\bibfnamefont {M.~M.}\ \bibnamefont
  {Wolf}}, \bibinfo {author} {\bibfnamefont {D.}~\bibnamefont {Perez-Garcia}},\
  and\ \bibinfo {author} {\bibfnamefont {J.~I.}\ \bibnamefont {Cirac}},\
  }\bibfield  {title} {\bibinfo {title} {Criticality, the area law, and the
  computational power of projected entangled pair states},\ }\href
  {https://doi.org/10.1103/PhysRevLett.96.220601} {\bibfield  {journal}
  {\bibinfo  {journal} {Phys. Rev. Lett.}\ }\textbf {\bibinfo {volume} {96}},\
  \bibinfo {pages} {220601} (\bibinfo {year} {2006})}\BibitemShut {NoStop}%
\bibitem [{\citenamefont {Valiant}(1979)}]{valiant1979complexity}%
  \BibitemOpen
  \bibfield  {author} {\bibinfo {author} {\bibfnamefont {L.~G.}\ \bibnamefont
  {Valiant}},\ }\bibfield  {title} {\bibinfo {title} {The complexity of
  computing the permanent},\ }\href@noop {} {\bibfield  {journal} {\bibinfo
  {journal} {Theoretical computer science}\ }\textbf {\bibinfo {volume} {8}},\
  \bibinfo {pages} {189} (\bibinfo {year} {1979})}\BibitemShut {NoStop}%
\bibitem [{\citenamefont {Shor}()}]{Shora}%
  \BibitemOpen
  \bibfield  {author} {\bibinfo {author} {\bibfnamefont {P.}~\bibnamefont
  {Shor}},\ }\bibfield  {title} {\bibinfo {title} {Algorithms for quantum
  computation: discrete logarithms and factoring},\ }in\ \href
  {https://doi.org/10.1109/sfcs.1994.365700} {\emph {\bibinfo {booktitle}
  {Proceedings 35th Annual Symposium on Foundations of Computer Science}}}\
  (\bibinfo  {publisher} {{IEEE} Comput. Soc. Press})\BibitemShut {NoStop}%
\bibitem [{\citenamefont {Shor}(1999)}]{Shor1999}%
  \BibitemOpen
  \bibfield  {author} {\bibinfo {author} {\bibfnamefont {P.~W.}\ \bibnamefont
  {Shor}},\ }\bibfield  {title} {\bibinfo {title} {Polynomial-time algorithms
  for prime factorization and discrete logarithms on a quantum computer},\
  }\href {https://doi.org/10.1137/s0036144598347011} {\bibfield  {journal}
  {\bibinfo  {journal} {{SIAM} Review}\ }\textbf {\bibinfo {volume} {41}},\
  \bibinfo {pages} {303} (\bibinfo {year} {1999})}\BibitemShut {NoStop}%
\bibitem [{\citenamefont {Grover}(1996)}]{Grover1996}%
  \BibitemOpen
  \bibfield  {author} {\bibinfo {author} {\bibfnamefont {L.~K.}\ \bibnamefont
  {Grover}},\ }\bibfield  {title} {\bibinfo {title} {A fast quantum mechanical
  algorithm for database search},\ }in\ \href
  {https://doi.org/10.1145/237814.237866} {\emph {\bibinfo {booktitle}
  {Proceedings of the twenty-eighth annual {ACM} symposium on Theory of
  computing - {STOC} {\textquotesingle}96}}}\ (\bibinfo  {publisher} {{ACM}
  Press},\ \bibinfo {year} {1996})\BibitemShut {NoStop}%
\bibitem [{\citenamefont {Grover}(1997)}]{Grover1997}%
  \BibitemOpen
  \bibfield  {author} {\bibinfo {author} {\bibfnamefont {L.~K.}\ \bibnamefont
  {Grover}},\ }\bibfield  {title} {\bibinfo {title} {Quantum mechanics helps in
  searching for a needle in a haystack},\ }\href
  {https://doi.org/10.1103/PhysRevLett.79.325} {\bibfield  {journal} {\bibinfo
  {journal} {Physical Review Letters}\ }\textbf {\bibinfo {volume} {79}},\
  \bibinfo {pages} {325} (\bibinfo {year} {1997})}\BibitemShut {NoStop}%
\bibitem [{\citenamefont {Farhi}\ \emph {et~al.}(2001)\citenamefont {Farhi},
  \citenamefont {Goldstone}, \citenamefont {Gutmann}, \citenamefont {Lapan},
  \citenamefont {Lundgren},\ and\ \citenamefont {Preda}}]{Farhi2001}%
  \BibitemOpen
  \bibfield  {author} {\bibinfo {author} {\bibfnamefont {E.}~\bibnamefont
  {Farhi}}, \bibinfo {author} {\bibfnamefont {J.}~\bibnamefont {Goldstone}},
  \bibinfo {author} {\bibfnamefont {S.}~\bibnamefont {Gutmann}}, \bibinfo
  {author} {\bibfnamefont {J.}~\bibnamefont {Lapan}}, \bibinfo {author}
  {\bibfnamefont {A.}~\bibnamefont {Lundgren}},\ and\ \bibinfo {author}
  {\bibfnamefont {D.}~\bibnamefont {Preda}},\ }\bibfield  {title} {\bibinfo
  {title} {A quantum adiabatic evolution algorithm applied to random instances
  of an {NP}-complete problem},\ }\href
  {https://doi.org/10.1126/science.1057726} {\bibfield  {journal} {\bibinfo
  {journal} {Science}\ }\textbf {\bibinfo {volume} {292}},\ \bibinfo {pages}
  {472} (\bibinfo {year} {2001})}\BibitemShut {NoStop}%
\bibitem [{\citenamefont {Farhi}\ \emph {et~al.}()\citenamefont {Farhi},
  \citenamefont {Goldstone},\ and\ \citenamefont {Gutmann}}]{Farhi2014}%
  \BibitemOpen
  \bibfield  {author} {\bibinfo {author} {\bibfnamefont {E.}~\bibnamefont
  {Farhi}}, \bibinfo {author} {\bibfnamefont {J.}~\bibnamefont {Goldstone}},\
  and\ \bibinfo {author} {\bibfnamefont {S.}~\bibnamefont {Gutmann}},\
  }\bibfield  {title} {\bibinfo {title} {A quantum approximate optimization
  algorithm},\ }\href {http://arxiv.org/abs/1411.4028v1} {\ }\Eprint
  {https://arxiv.org/abs/arXiv:1411.4028v1} {arXiv:1411.4028v1} \BibitemShut
  {NoStop}%
\bibitem [{\citenamefont {Brooke}(1999)}]{Brooke1999}%
  \BibitemOpen
  \bibfield  {author} {\bibinfo {author} {\bibfnamefont {J.}~\bibnamefont
  {Brooke}},\ }\bibfield  {title} {\bibinfo {title} {Quantum annealing of a
  disordered magnet},\ }\href {https://doi.org/10.1126/science.284.5415.779}
  {\bibfield  {journal} {\bibinfo  {journal} {Science}\ }\textbf {\bibinfo
  {volume} {284}},\ \bibinfo {pages} {779} (\bibinfo {year}
  {1999})}\BibitemShut {NoStop}%
\bibitem [{\citenamefont {Santoro}(2002)}]{Santoro2002}%
  \BibitemOpen
  \bibfield  {author} {\bibinfo {author} {\bibfnamefont {G.~E.}\ \bibnamefont
  {Santoro}},\ }\bibfield  {title} {\bibinfo {title} {Theory of quantum
  annealing of an ising spin glass},\ }\href
  {https://doi.org/10.1126/science.1068774} {\bibfield  {journal} {\bibinfo
  {journal} {Science}\ }\textbf {\bibinfo {volume} {295}},\ \bibinfo {pages}
  {2427} (\bibinfo {year} {2002})}\BibitemShut {NoStop}%
\bibitem [{\citenamefont {Grimsley}\ \emph {et~al.}(2019)\citenamefont
  {Grimsley}, \citenamefont {Economou}, \citenamefont {Barnes},\ and\
  \citenamefont {Mayhall}}]{Grimsley2019}%
  \BibitemOpen
  \bibfield  {author} {\bibinfo {author} {\bibfnamefont {H.~R.}\ \bibnamefont
  {Grimsley}}, \bibinfo {author} {\bibfnamefont {S.~E.}\ \bibnamefont
  {Economou}}, \bibinfo {author} {\bibfnamefont {E.}~\bibnamefont {Barnes}},\
  and\ \bibinfo {author} {\bibfnamefont {N.~J.}\ \bibnamefont {Mayhall}},\
  }\bibfield  {title} {\bibinfo {title} {An adaptive variational algorithm for
  exact molecular simulations on a quantum computer},\ }\bibfield  {journal}
  {\bibinfo  {journal} {Nature Communications}\ }\textbf {\bibinfo {volume}
  {10}},\ \href {https://doi.org/10.1038/s41467-019-10988-2}
  {10.1038/s41467-019-10988-2} (\bibinfo {year} {2019})\BibitemShut {NoStop}%
\bibitem [{\citenamefont {Schuch}\ \emph {et~al.}(2010)\citenamefont {Schuch},
  \citenamefont {Cirac},\ and\ \citenamefont
  {Pérez-García}}]{SCHUCH20102153}%
  \BibitemOpen
  \bibfield  {author} {\bibinfo {author} {\bibfnamefont {N.}~\bibnamefont
  {Schuch}}, \bibinfo {author} {\bibfnamefont {I.}~\bibnamefont {Cirac}},\ and\
  \bibinfo {author} {\bibfnamefont {D.}~\bibnamefont {Pérez-García}},\
  }\bibfield  {title} {\bibinfo {title} {Peps as ground states: Degeneracy and
  topology},\ }\href
  {https://doi.org/https://doi.org/10.1016/j.aop.2010.05.008} {\bibfield
  {journal} {\bibinfo  {journal} {Annals of Physics}\ }\textbf {\bibinfo
  {volume} {325}},\ \bibinfo {pages} {2153 } (\bibinfo {year}
  {2010})}\BibitemShut {NoStop}%
\bibitem [{\citenamefont {Sch\"on}\ \emph {et~al.}(2005)\citenamefont
  {Sch\"on}, \citenamefont {Solano}, \citenamefont {Verstraete}, \citenamefont
  {Cirac},\ and\ \citenamefont {Wolf}}]{PhysRevLett.95.110503}%
  \BibitemOpen
  \bibfield  {author} {\bibinfo {author} {\bibfnamefont {C.}~\bibnamefont
  {Sch\"on}}, \bibinfo {author} {\bibfnamefont {E.}~\bibnamefont {Solano}},
  \bibinfo {author} {\bibfnamefont {F.}~\bibnamefont {Verstraete}}, \bibinfo
  {author} {\bibfnamefont {J.~I.}\ \bibnamefont {Cirac}},\ and\ \bibinfo
  {author} {\bibfnamefont {M.~M.}\ \bibnamefont {Wolf}},\ }\bibfield  {title}
  {\bibinfo {title} {Sequential generation of entangled multiqubit states},\
  }\href {https://doi.org/10.1103/PhysRevLett.95.110503} {\bibfield  {journal}
  {\bibinfo  {journal} {Phys. Rev. Lett.}\ }\textbf {\bibinfo {volume} {95}},\
  \bibinfo {pages} {110503} (\bibinfo {year} {2005})}\BibitemShut {NoStop}%
\bibitem [{\citenamefont {Cramer}\ \emph {et~al.}(2010)\citenamefont {Cramer},
  \citenamefont {Plenio}, \citenamefont {Flammia}, \citenamefont {Somma},
  \citenamefont {Gross}, \citenamefont {Bartlett}, \citenamefont
  {Landon-Cardinal}, \citenamefont {Poulin},\ and\ \citenamefont
  {Liu}}]{cramer2010efficient}%
  \BibitemOpen
  \bibfield  {author} {\bibinfo {author} {\bibfnamefont {M.}~\bibnamefont
  {Cramer}}, \bibinfo {author} {\bibfnamefont {M.~B.}\ \bibnamefont {Plenio}},
  \bibinfo {author} {\bibfnamefont {S.~T.}\ \bibnamefont {Flammia}}, \bibinfo
  {author} {\bibfnamefont {R.}~\bibnamefont {Somma}}, \bibinfo {author}
  {\bibfnamefont {D.}~\bibnamefont {Gross}}, \bibinfo {author} {\bibfnamefont
  {S.~D.}\ \bibnamefont {Bartlett}}, \bibinfo {author} {\bibfnamefont
  {O.}~\bibnamefont {Landon-Cardinal}}, \bibinfo {author} {\bibfnamefont
  {D.}~\bibnamefont {Poulin}},\ and\ \bibinfo {author} {\bibfnamefont {Y.-K.}\
  \bibnamefont {Liu}},\ }\bibfield  {title} {\bibinfo {title} {Efficient
  quantum state tomography},\ }\href@noop {} {\bibfield  {journal} {\bibinfo
  {journal} {Nature communications}\ }\textbf {\bibinfo {volume} {1}},\
  \bibinfo {pages} {1} (\bibinfo {year} {2010})}\BibitemShut {NoStop}%
\bibitem [{\citenamefont {Eichler}\ \emph {et~al.}(2015)\citenamefont
  {Eichler}, \citenamefont {Mlynek}, \citenamefont {Butscher}, \citenamefont
  {Kurpiers}, \citenamefont {Hammerer}, \citenamefont {Osborne},\ and\
  \citenamefont {Wallraff}}]{PhysRevX.5.041044}%
  \BibitemOpen
  \bibfield  {author} {\bibinfo {author} {\bibfnamefont {C.}~\bibnamefont
  {Eichler}}, \bibinfo {author} {\bibfnamefont {J.}~\bibnamefont {Mlynek}},
  \bibinfo {author} {\bibfnamefont {J.}~\bibnamefont {Butscher}}, \bibinfo
  {author} {\bibfnamefont {P.}~\bibnamefont {Kurpiers}}, \bibinfo {author}
  {\bibfnamefont {K.}~\bibnamefont {Hammerer}}, \bibinfo {author}
  {\bibfnamefont {T.~J.}\ \bibnamefont {Osborne}},\ and\ \bibinfo {author}
  {\bibfnamefont {A.}~\bibnamefont {Wallraff}},\ }\bibfield  {title} {\bibinfo
  {title} {Exploring interacting quantum many-body systems by experimentally
  creating continuous matrix product states in superconducting circuits},\
  }\href {https://doi.org/10.1103/PhysRevX.5.041044} {\bibfield  {journal}
  {\bibinfo  {journal} {Phys. Rev. X}\ }\textbf {\bibinfo {volume} {5}},\
  \bibinfo {pages} {041044} (\bibinfo {year} {2015})}\BibitemShut {NoStop}%
\bibitem [{\citenamefont {Liu}\ \emph {et~al.}(2019)\citenamefont {Liu},
  \citenamefont {Zhang}, \citenamefont {Wan},\ and\ \citenamefont
  {Wang}}]{PhysRevResearch.1.023025}%
  \BibitemOpen
  \bibfield  {author} {\bibinfo {author} {\bibfnamefont {J.-G.}\ \bibnamefont
  {Liu}}, \bibinfo {author} {\bibfnamefont {Y.-H.}\ \bibnamefont {Zhang}},
  \bibinfo {author} {\bibfnamefont {Y.}~\bibnamefont {Wan}},\ and\ \bibinfo
  {author} {\bibfnamefont {L.}~\bibnamefont {Wang}},\ }\bibfield  {title}
  {\bibinfo {title} {Variational quantum eigensolver with fewer qubits},\
  }\href {https://doi.org/10.1103/PhysRevResearch.1.023025} {\bibfield
  {journal} {\bibinfo  {journal} {Phys. Rev. Research}\ }\textbf {\bibinfo
  {volume} {1}},\ \bibinfo {pages} {023025} (\bibinfo {year}
  {2019})}\BibitemShut {NoStop}%
\bibitem [{\citenamefont {{Ran}}(2019)}]{2019arXiv190807958R}%
  \BibitemOpen
  \bibfield  {author} {\bibinfo {author} {\bibfnamefont {S.-J.}\ \bibnamefont
  {{Ran}}},\ }\bibfield  {title} {\bibinfo {title} {{Efficient Encoding of
  Matrix Product States into Quantum Circuits of One- and Two-Qubit Gates}},\
  }\href {https://ui.adsabs.harvard.edu/abs/2019arXiv190807958R} {\ ,\ \bibinfo
  {eid} {arXiv:1908.07958} (\bibinfo {year} {2019})}\BibitemShut {NoStop}%
\bibitem [{\citenamefont {Schwarz}\ \emph {et~al.}(2013)\citenamefont
  {Schwarz}, \citenamefont {Temme}, \citenamefont {Verstraete}, \citenamefont
  {Perez-Garcia},\ and\ \citenamefont {Cubitt}}]{PhysRevA.88.032321}%
  \BibitemOpen
  \bibfield  {author} {\bibinfo {author} {\bibfnamefont {M.}~\bibnamefont
  {Schwarz}}, \bibinfo {author} {\bibfnamefont {K.}~\bibnamefont {Temme}},
  \bibinfo {author} {\bibfnamefont {F.}~\bibnamefont {Verstraete}}, \bibinfo
  {author} {\bibfnamefont {D.}~\bibnamefont {Perez-Garcia}},\ and\ \bibinfo
  {author} {\bibfnamefont {T.~S.}\ \bibnamefont {Cubitt}},\ }\bibfield  {title}
  {\bibinfo {title} {Preparing topological projected entangled pair states on a
  quantum computer},\ }\href {https://doi.org/10.1103/PhysRevA.88.032321}
  {\bibfield  {journal} {\bibinfo  {journal} {Phys. Rev. A}\ }\textbf {\bibinfo
  {volume} {88}},\ \bibinfo {pages} {032321} (\bibinfo {year}
  {2013})}\BibitemShut {NoStop}%
\bibitem [{\citenamefont {Schwarz}\ \emph {et~al.}(2012)\citenamefont
  {Schwarz}, \citenamefont {Temme},\ and\ \citenamefont
  {Verstraete}}]{PhysRevLett.108.110502}%
  \BibitemOpen
  \bibfield  {author} {\bibinfo {author} {\bibfnamefont {M.}~\bibnamefont
  {Schwarz}}, \bibinfo {author} {\bibfnamefont {K.}~\bibnamefont {Temme}},\
  and\ \bibinfo {author} {\bibfnamefont {F.}~\bibnamefont {Verstraete}},\
  }\bibfield  {title} {\bibinfo {title} {Preparing projected entangled pair
  states on a quantum computer},\ }\href
  {https://doi.org/10.1103/PhysRevLett.108.110502} {\bibfield  {journal}
  {\bibinfo  {journal} {Phys. Rev. Lett.}\ }\textbf {\bibinfo {volume} {108}},\
  \bibinfo {pages} {110502} (\bibinfo {year} {2012})}\BibitemShut {NoStop}%
\bibitem [{\citenamefont {Arad}\ and\ \citenamefont {Landau}()}]{Arad2008}%
  \BibitemOpen
  \bibfield  {author} {\bibinfo {author} {\bibfnamefont {I.}~\bibnamefont
  {Arad}}\ and\ \bibinfo {author} {\bibfnamefont {Z.}~\bibnamefont {Landau}},\
  }\bibfield  {title} {\bibinfo {title} {Quantum computation and the evaluation
  of tensor networks},\ }\href {http://arxiv.org/abs/0805.0040v3} {\ }\Eprint
  {https://arxiv.org/abs/arXiv:0805.0040v3} {arXiv:0805.0040v3} \BibitemShut
  {NoStop}%
\bibitem [{\citenamefont {Mazzola}\ \emph {et~al.}(2019)\citenamefont
  {Mazzola}, \citenamefont {Ollitrault}, \citenamefont {Barkoutsos},\ and\
  \citenamefont {Tavernelli}}]{Mazzola2019}%
  \BibitemOpen
  \bibfield  {author} {\bibinfo {author} {\bibfnamefont {G.}~\bibnamefont
  {Mazzola}}, \bibinfo {author} {\bibfnamefont {P.~J.}\ \bibnamefont
  {Ollitrault}}, \bibinfo {author} {\bibfnamefont {P.~K.}\ \bibnamefont
  {Barkoutsos}},\ and\ \bibinfo {author} {\bibfnamefont {I.}~\bibnamefont
  {Tavernelli}},\ }\bibfield  {title} {\bibinfo {title} {Nonunitary operations
  for ground-state calculations in near-term quantum computers},\ }\bibfield
  {journal} {\bibinfo  {journal} {Physical Review Letters}\ }\textbf {\bibinfo
  {volume} {123}},\ \href {https://doi.org/10.1103/PhysRevLett.123.130501}
  {10.1103/PhysRevLett.123.130501} (\bibinfo {year} {2019})\BibitemShut
  {NoStop}%
\bibitem [{\citenamefont {Seki}\ \emph {et~al.}()\citenamefont {Seki},
  \citenamefont {Shirakawa},\ and\ \citenamefont {Yunoki}}]{Seki2019}%
  \BibitemOpen
  \bibfield  {author} {\bibinfo {author} {\bibfnamefont {K.}~\bibnamefont
  {Seki}}, \bibinfo {author} {\bibfnamefont {T.}~\bibnamefont {Shirakawa}},\
  and\ \bibinfo {author} {\bibfnamefont {S.}~\bibnamefont {Yunoki}},\
  }\bibfield  {title} {\bibinfo {title} {Symmetry-adapted variational quantum
  eigensolver},\ }\href@noop {} {\ }\Eprint
  {https://arxiv.org/abs/http://arxiv.org/abs/1912.13146v1}
  {http://arxiv.org/abs/1912.13146v1} \BibitemShut {NoStop}%
\bibitem [{\citenamefont {Motta}\ \emph {et~al.}(2019)\citenamefont {Motta},
  \citenamefont {Sun}, \citenamefont {Tan}, \citenamefont {O'Rourke},
  \citenamefont {Ye}, \citenamefont {Minnich}, \citenamefont {Brand{\~{a}}o},\
  and\ \citenamefont {Chan}}]{Motta2019}%
  \BibitemOpen
  \bibfield  {author} {\bibinfo {author} {\bibfnamefont {M.}~\bibnamefont
  {Motta}}, \bibinfo {author} {\bibfnamefont {C.}~\bibnamefont {Sun}}, \bibinfo
  {author} {\bibfnamefont {A.~T.~K.}\ \bibnamefont {Tan}}, \bibinfo {author}
  {\bibfnamefont {M.~J.}\ \bibnamefont {O'Rourke}}, \bibinfo {author}
  {\bibfnamefont {E.}~\bibnamefont {Ye}}, \bibinfo {author} {\bibfnamefont
  {A.~J.}\ \bibnamefont {Minnich}}, \bibinfo {author} {\bibfnamefont {F.~G.
  S.~L.}\ \bibnamefont {Brand{\~{a}}o}},\ and\ \bibinfo {author} {\bibfnamefont
  {G.~K.-L.}\ \bibnamefont {Chan}},\ }\bibfield  {title} {\bibinfo {title}
  {Determining eigenstates and thermal states on a quantum computer using
  quantum imaginary time evolution},\ }\bibfield  {journal} {\bibinfo
  {journal} {Nature Physics}\ }\href
  {https://doi.org/10.1038/s41567-019-0704-4} {10.1038/s41567-019-0704-4}
  (\bibinfo {year} {2019})\BibitemShut {NoStop}%
\bibitem [{\citenamefont {Uhlmann}(1976)}]{uhlmann1976transition}%
  \BibitemOpen
  \bibfield  {author} {\bibinfo {author} {\bibfnamefont {A.}~\bibnamefont
  {Uhlmann}},\ }\bibfield  {title} {\bibinfo {title} {The ``transition
  probability" in the state space of a*-algebra},\ }\href@noop {} {\bibfield
  {journal} {\bibinfo  {journal} {Reports on Mathematical Physics}\ }\textbf
  {\bibinfo {volume} {9}},\ \bibinfo {pages} {273} (\bibinfo {year}
  {1976})}\BibitemShut {NoStop}%
\bibitem [{\citenamefont {Wang}\ \emph {et~al.}(2010)\citenamefont {Wang},
  \citenamefont {Wu}, \citenamefont {xi~Liu},\ and\ \citenamefont
  {Nori}}]{Wang2010}%
  \BibitemOpen
  \bibfield  {author} {\bibinfo {author} {\bibfnamefont {H.}~\bibnamefont
  {Wang}}, \bibinfo {author} {\bibfnamefont {L.-A.}\ \bibnamefont {Wu}},
  \bibinfo {author} {\bibfnamefont {Y.}~\bibnamefont {xi~Liu}},\ and\ \bibinfo
  {author} {\bibfnamefont {F.}~\bibnamefont {Nori}},\ }\bibfield  {title}
  {\bibinfo {title} {Measurement-based quantum phase estimation algorithm for
  finding eigenvalues of non-unitary matrices},\ }\bibfield  {journal}
  {\bibinfo  {journal} {Physical Review A}\ }\textbf {\bibinfo {volume} {82}},\
  \href {https://doi.org/10.1103/physreva.82.062303}
  {10.1103/physreva.82.062303} (\bibinfo {year} {2010})\BibitemShut {NoStop}%
\bibitem [{\citenamefont {Kong}\ \emph {et~al.}(2019)\citenamefont {Kong},
  \citenamefont {Wei}, \citenamefont {Wen}, \citenamefont {Xin},\ and\
  \citenamefont {Long}}]{Kong2019}%
  \BibitemOpen
  \bibfield  {author} {\bibinfo {author} {\bibfnamefont {X.}~\bibnamefont
  {Kong}}, \bibinfo {author} {\bibfnamefont {S.}~\bibnamefont {Wei}}, \bibinfo
  {author} {\bibfnamefont {J.}~\bibnamefont {Wen}}, \bibinfo {author}
  {\bibfnamefont {T.}~\bibnamefont {Xin}},\ and\ \bibinfo {author}
  {\bibfnamefont {G.-L.}\ \bibnamefont {Long}},\ }\bibfield  {title} {\bibinfo
  {title} {Experimental simulation of shift operators in a quantum processor},\
  }\bibfield  {journal} {\bibinfo  {journal} {Physical Review A}\ }\textbf
  {\bibinfo {volume} {99}},\ \href {https://doi.org/10.1103/physreva.99.042328}
  {10.1103/physreva.99.042328} (\bibinfo {year} {2019})\BibitemShut {NoStop}%
\bibitem [{\citenamefont {Williams}(2004)}]{Williams2004}%
  \BibitemOpen
  \bibfield  {author} {\bibinfo {author} {\bibfnamefont {C.~P.}\ \bibnamefont
  {Williams}},\ }\bibfield  {title} {\bibinfo {title} {Probabilistic nonunitary
  quantum computing},\ }in\ \href {https://doi.org/10.1117/12.542413} {\emph
  {\bibinfo {booktitle} {Quantum Information and Computation {II}}}},\ \bibinfo
  {editor} {edited by\ \bibinfo {editor} {\bibfnamefont {E.}~\bibnamefont
  {Donkor}}, \bibinfo {editor} {\bibfnamefont {A.~R.}\ \bibnamefont {Pirich}},\
  and\ \bibinfo {editor} {\bibfnamefont {H.~E.}\ \bibnamefont {Brandt}}}\
  (\bibinfo  {publisher} {{SPIE}},\ \bibinfo {year} {2004})\BibitemShut
  {NoStop}%
\bibitem [{\citenamefont {TERASHIMA}\ and\ \citenamefont
  {UEDA}(2005)}]{TERASHIMA2005}%
  \BibitemOpen
  \bibfield  {author} {\bibinfo {author} {\bibfnamefont {H.}~\bibnamefont
  {TERASHIMA}}\ and\ \bibinfo {author} {\bibfnamefont {M.}~\bibnamefont
  {UEDA}},\ }\bibfield  {title} {\bibinfo {title} {{NONUNITARY} {QUANTUM}
  {CIRCUIT}},\ }\href {https://doi.org/10.1142/s0219749905001456} {\bibfield
  {journal} {\bibinfo  {journal} {International Journal of Quantum
  Information}\ }\textbf {\bibinfo {volume} {03}},\ \bibinfo {pages} {633}
  (\bibinfo {year} {2005})}\BibitemShut {NoStop}%
\bibitem [{\citenamefont {Barnett}\ \emph {et~al.}(1998)\citenamefont
  {Barnett}, \citenamefont {Jeffers}, \citenamefont {Gatti},\ and\
  \citenamefont {Loudon}}]{Barnett1998}%
  \BibitemOpen
  \bibfield  {author} {\bibinfo {author} {\bibfnamefont {S.~M.}\ \bibnamefont
  {Barnett}}, \bibinfo {author} {\bibfnamefont {J.}~\bibnamefont {Jeffers}},
  \bibinfo {author} {\bibfnamefont {A.}~\bibnamefont {Gatti}},\ and\ \bibinfo
  {author} {\bibfnamefont {R.}~\bibnamefont {Loudon}},\ }\bibfield  {title}
  {\bibinfo {title} {Quantum optics of lossy beam splitters},\ }\href
  {https://doi.org/10.1103/physreva.57.2134} {\bibfield  {journal} {\bibinfo
  {journal} {Physical Review A}\ }\textbf {\bibinfo {volume} {57}},\ \bibinfo
  {pages} {2134} (\bibinfo {year} {1998})}\BibitemShut {NoStop}%
\bibitem [{\citenamefont {Klyshko}(1989)}]{klyshko1989nonunitary}%
  \BibitemOpen
  \bibfield  {author} {\bibinfo {author} {\bibfnamefont {D.}~\bibnamefont
  {Klyshko}},\ }\bibfield  {title} {\bibinfo {title} {Nonunitary
  transformations in quantum optics},\ }\href@noop {} {\bibfield  {journal}
  {\bibinfo  {journal} {Physics Letters A}\ }\textbf {\bibinfo {volume}
  {137}},\ \bibinfo {pages} {334} (\bibinfo {year} {1989})}\BibitemShut
  {NoStop}%
\bibitem [{\citenamefont {Roger}\ \emph {et~al.}(2015)\citenamefont {Roger},
  \citenamefont {Vezzoli}, \citenamefont {Bolduc}, \citenamefont {Valente},
  \citenamefont {Heitz}, \citenamefont {Jeffers}, \citenamefont {Soci},
  \citenamefont {Leach}, \citenamefont {Couteau}, \citenamefont {Zheludev},\
  and\ \citenamefont {Faccio}}]{Roger2015}%
  \BibitemOpen
  \bibfield  {author} {\bibinfo {author} {\bibfnamefont {T.}~\bibnamefont
  {Roger}}, \bibinfo {author} {\bibfnamefont {S.}~\bibnamefont {Vezzoli}},
  \bibinfo {author} {\bibfnamefont {E.}~\bibnamefont {Bolduc}}, \bibinfo
  {author} {\bibfnamefont {J.}~\bibnamefont {Valente}}, \bibinfo {author}
  {\bibfnamefont {J.~J.~F.}\ \bibnamefont {Heitz}}, \bibinfo {author}
  {\bibfnamefont {J.}~\bibnamefont {Jeffers}}, \bibinfo {author} {\bibfnamefont
  {C.}~\bibnamefont {Soci}}, \bibinfo {author} {\bibfnamefont {J.}~\bibnamefont
  {Leach}}, \bibinfo {author} {\bibfnamefont {C.}~\bibnamefont {Couteau}},
  \bibinfo {author} {\bibfnamefont {N.~I.}\ \bibnamefont {Zheludev}},\ and\
  \bibinfo {author} {\bibfnamefont {D.}~\bibnamefont {Faccio}},\ }\bibfield
  {title} {\bibinfo {title} {Coherent perfect absorption in deeply
  subwavelength films in the single-photon regime},\ }\bibfield  {journal}
  {\bibinfo  {journal} {Nature Communications}\ }\textbf {\bibinfo {volume}
  {6}},\ \href {https://doi.org/10.1038/ncomms8031} {10.1038/ncomms8031}
  (\bibinfo {year} {2015})\BibitemShut {NoStop}%
\bibitem [{\citenamefont {Tischler}\ \emph {et~al.}(2018)\citenamefont
  {Tischler}, \citenamefont {Rockstuhl},\ and\ \citenamefont
  {S{\l}owik}}]{Tischler2018}%
  \BibitemOpen
  \bibfield  {author} {\bibinfo {author} {\bibfnamefont {N.}~\bibnamefont
  {Tischler}}, \bibinfo {author} {\bibfnamefont {C.}~\bibnamefont
  {Rockstuhl}},\ and\ \bibinfo {author} {\bibfnamefont {K.}~\bibnamefont
  {S{\l}owik}},\ }\bibfield  {title} {\bibinfo {title} {Quantum optical
  realization of arbitrary linear transformations allowing for loss and gain},\
  }\bibfield  {journal} {\bibinfo  {journal} {Physical Review X}\ }\textbf
  {\bibinfo {volume} {8}},\ \href {https://doi.org/10.1103/physrevx.8.021017}
  {10.1103/physrevx.8.021017} (\bibinfo {year} {2018})\BibitemShut {NoStop}%
\bibitem [{\citenamefont {Childs}\ and\ \citenamefont {Wiebe}()}]{Childs2012}%
  \BibitemOpen
  \bibfield  {author} {\bibinfo {author} {\bibfnamefont {A.~M.}\ \bibnamefont
  {Childs}}\ and\ \bibinfo {author} {\bibfnamefont {N.}~\bibnamefont {Wiebe}},\
  }\bibfield  {title} {\bibinfo {title} {Hamiltonian simulation using linear
  combinations of unitary operations}\ }\href
  {https://doi.org/10.26421/QIC12.11-12} {10.26421/QIC12.11-12},\ \Eprint
  {https://arxiv.org/abs/http://arxiv.org/abs/1202.5822v1}
  {http://arxiv.org/abs/1202.5822v1} \BibitemShut {NoStop}%
\bibitem [{\citenamefont {Berry}\ \emph {et~al.}(2015)\citenamefont {Berry},
  \citenamefont {Childs}, \citenamefont {Cleve}, \citenamefont {Kothari},\ and\
  \citenamefont {Somma}}]{Berry2015}%
  \BibitemOpen
  \bibfield  {author} {\bibinfo {author} {\bibfnamefont {D.~W.}\ \bibnamefont
  {Berry}}, \bibinfo {author} {\bibfnamefont {A.~M.}\ \bibnamefont {Childs}},
  \bibinfo {author} {\bibfnamefont {R.}~\bibnamefont {Cleve}}, \bibinfo
  {author} {\bibfnamefont {R.}~\bibnamefont {Kothari}},\ and\ \bibinfo {author}
  {\bibfnamefont {R.~D.}\ \bibnamefont {Somma}},\ }\bibfield  {title} {\bibinfo
  {title} {Simulating hamiltonian dynamics with a truncated taylor series},\
  }\bibfield  {journal} {\bibinfo  {journal} {Physical Review Letters}\
  }\textbf {\bibinfo {volume} {114}},\ \href
  {https://doi.org/10.1103/PhysRevLett.114.090502}
  {10.1103/PhysRevLett.114.090502} (\bibinfo {year} {2015})\BibitemShut
  {NoStop}%
\bibitem [{\citenamefont {McArdle}\ \emph {et~al.}(2019)\citenamefont
  {McArdle}, \citenamefont {Jones}, \citenamefont {Endo}, \citenamefont {Li},
  \citenamefont {Benjamin},\ and\ \citenamefont {Yuan}}]{McArdle2019}%
  \BibitemOpen
  \bibfield  {author} {\bibinfo {author} {\bibfnamefont {S.}~\bibnamefont
  {McArdle}}, \bibinfo {author} {\bibfnamefont {T.}~\bibnamefont {Jones}},
  \bibinfo {author} {\bibfnamefont {S.}~\bibnamefont {Endo}}, \bibinfo {author}
  {\bibfnamefont {Y.}~\bibnamefont {Li}}, \bibinfo {author} {\bibfnamefont
  {S.~C.}\ \bibnamefont {Benjamin}},\ and\ \bibinfo {author} {\bibfnamefont
  {X.}~\bibnamefont {Yuan}},\ }\bibfield  {title} {\bibinfo {title}
  {Variational ansatz-based quantum simulation of imaginary time evolution},\
  }\bibfield  {journal} {\bibinfo  {journal} {npj Quantum Information}\
  }\textbf {\bibinfo {volume} {5}},\ \href
  {https://doi.org/10.1038/s41534-019-0187-2} {10.1038/s41534-019-0187-2}
  (\bibinfo {year} {2019})\BibitemShut {NoStop}%
\bibitem [{\citenamefont {Vidal}(2004)}]{Vidal2004a}%
  \BibitemOpen
  \bibfield  {author} {\bibinfo {author} {\bibfnamefont {G.}~\bibnamefont
  {Vidal}},\ }\bibfield  {title} {\bibinfo {title} {Efficient simulation of
  one-dimensional quantum many-body systems},\ }\bibfield  {journal} {\bibinfo
  {journal} {Physical Review Letters}\ }\textbf {\bibinfo {volume} {93}},\
  \href {https://doi.org/10.1103/physrevlett.93.040502}
  {10.1103/physrevlett.93.040502} (\bibinfo {year} {2004})\BibitemShut
  {NoStop}%
\bibitem [{\citenamefont {Luo}\ \emph {et~al.}(2019)\citenamefont {Luo},
  \citenamefont {Liu}, \citenamefont {Zhang},\ and\ \citenamefont
  {Wang}}]{YaoFramework2019}%
  \BibitemOpen
  \bibfield  {author} {\bibinfo {author} {\bibfnamefont {X.-Z.}\ \bibnamefont
  {Luo}}, \bibinfo {author} {\bibfnamefont {J.-G.}\ \bibnamefont {Liu}},
  \bibinfo {author} {\bibfnamefont {P.}~\bibnamefont {Zhang}},\ and\ \bibinfo
  {author} {\bibfnamefont {L.}~\bibnamefont {Wang}},\ }\bibfield  {title}
  {\bibinfo {title} {Yao.jl: Extensible, efficient framework for quantum
  algorithm design},\ }\href@noop {} {\bibfield  {journal} {\bibinfo  {journal}
  {arXiv preprint arXiv:1912.10877}\ } (\bibinfo {year} {2019})}\BibitemShut
  {NoStop}%
\bibitem [{\citenamefont {Garey}\ \emph {et~al.}(1974)\citenamefont {Garey},
  \citenamefont {Johnson},\ and\ \citenamefont {Stockmeyer}}]{Garey1974}%
  \BibitemOpen
  \bibfield  {author} {\bibinfo {author} {\bibfnamefont {M.~R.}\ \bibnamefont
  {Garey}}, \bibinfo {author} {\bibfnamefont {D.~S.}\ \bibnamefont {Johnson}},\
  and\ \bibinfo {author} {\bibfnamefont {L.}~\bibnamefont {Stockmeyer}},\
  }\bibfield  {title} {\bibinfo {title} {Some simplified {NP}-complete
  problems},\ }in\ \href {https://doi.org/10.1145/800119.803884} {\emph
  {\bibinfo {booktitle} {Proceedings of the sixth annual {ACM} symposium on
  Theory of computing - {STOC} {\textquotesingle}74}}}\ (\bibinfo  {publisher}
  {{ACM} Press},\ \bibinfo {year} {1974})\BibitemShut {NoStop}%
\bibitem [{\citenamefont {Papadimitriou}\ and\ \citenamefont
  {Yannakakis}(1991)}]{Papadimitriou1991}%
  \BibitemOpen
  \bibfield  {author} {\bibinfo {author} {\bibfnamefont {C.~H.}\ \bibnamefont
  {Papadimitriou}}\ and\ \bibinfo {author} {\bibfnamefont {M.}~\bibnamefont
  {Yannakakis}},\ }\bibfield  {title} {\bibinfo {title} {Optimization,
  approximation, and complexity classes},\ }\href
  {https://doi.org/10.1016/0022-0000(91)90023-x} {\bibfield  {journal}
  {\bibinfo  {journal} {Journal of Computer and System Sciences}\ }\textbf
  {\bibinfo {volume} {43}},\ \bibinfo {pages} {425} (\bibinfo {year}
  {1991})}\BibitemShut {NoStop}%
\bibitem [{\citenamefont {Barahona}\ \emph {et~al.}(1988)\citenamefont
  {Barahona}, \citenamefont {Grötschel}, \citenamefont {Jünger},\ and\
  \citenamefont {Reinelt}}]{Barahona1988}%
  \BibitemOpen
  \bibfield  {author} {\bibinfo {author} {\bibfnamefont {F.}~\bibnamefont
  {Barahona}}, \bibinfo {author} {\bibfnamefont {M.}~\bibnamefont
  {Grötschel}}, \bibinfo {author} {\bibfnamefont {M.}~\bibnamefont
  {Jünger}},\ and\ \bibinfo {author} {\bibfnamefont {G.}~\bibnamefont
  {Reinelt}},\ }\bibfield  {title} {\bibinfo {title} {An application of
  combinatorial optimization to statistical physics and circuit layout
  design},\ }\href {https://doi.org/10.1287/opre.36.3.493} {\bibfield
  {journal} {\bibinfo  {journal} {Operations Research}\ }\textbf {\bibinfo
  {volume} {36}},\ \bibinfo {pages} {493} (\bibinfo {year} {1988})}\BibitemShut
  {NoStop}%
\bibitem [{\citenamefont {Kirkpatrick}\ \emph {et~al.}(1983)\citenamefont
  {Kirkpatrick}, \citenamefont {Gelatt},\ and\ \citenamefont
  {Vecchi}}]{kirkpatrick1983optimization}%
  \BibitemOpen
  \bibfield  {author} {\bibinfo {author} {\bibfnamefont {S.}~\bibnamefont
  {Kirkpatrick}}, \bibinfo {author} {\bibfnamefont {C.~D.}\ \bibnamefont
  {Gelatt}},\ and\ \bibinfo {author} {\bibfnamefont {M.~P.}\ \bibnamefont
  {Vecchi}},\ }\bibfield  {title} {\bibinfo {title} {Optimization by simulated
  annealing},\ }\href@noop {} {\bibfield  {journal} {\bibinfo  {journal}
  {science}\ }\textbf {\bibinfo {volume} {220}},\ \bibinfo {pages} {671}
  (\bibinfo {year} {1983})}\BibitemShut {NoStop}%
\bibitem [{\citenamefont {Peng}\ \emph {et~al.}(2014)\citenamefont {Peng},
  \citenamefont {Luo}, \citenamefont {Zheng}, \citenamefont {Kou},
  \citenamefont {Suter},\ and\ \citenamefont {Du}}]{Peng2014}%
  \BibitemOpen
  \bibfield  {author} {\bibinfo {author} {\bibfnamefont {X.}~\bibnamefont
  {Peng}}, \bibinfo {author} {\bibfnamefont {Z.}~\bibnamefont {Luo}}, \bibinfo
  {author} {\bibfnamefont {W.}~\bibnamefont {Zheng}}, \bibinfo {author}
  {\bibfnamefont {S.}~\bibnamefont {Kou}}, \bibinfo {author} {\bibfnamefont
  {D.}~\bibnamefont {Suter}},\ and\ \bibinfo {author} {\bibfnamefont
  {J.}~\bibnamefont {Du}},\ }\bibfield  {title} {\bibinfo {title} {Experimental
  implementation of adiabatic passage between different topological orders},\
  }\href {https://doi.org/10.1103/PhysRevLett.113.080404} {\bibfield  {journal}
  {\bibinfo  {journal} {Phys. Rev. Lett.}\ }\textbf {\bibinfo {volume} {113}},\
  \bibinfo {pages} {080404} (\bibinfo {year} {2014})}\BibitemShut {NoStop}%
\bibitem [{\citenamefont {Tseng}\ \emph {et~al.}(1999)\citenamefont {Tseng},
  \citenamefont {Somaroo}, \citenamefont {Sharf}, \citenamefont {Knill},
  \citenamefont {Laflamme}, \citenamefont {Havel},\ and\ \citenamefont
  {Cory}}]{Tseng1999}%
  \BibitemOpen
  \bibfield  {author} {\bibinfo {author} {\bibfnamefont {C.~H.}\ \bibnamefont
  {Tseng}}, \bibinfo {author} {\bibfnamefont {S.}~\bibnamefont {Somaroo}},
  \bibinfo {author} {\bibfnamefont {Y.}~\bibnamefont {Sharf}}, \bibinfo
  {author} {\bibfnamefont {E.}~\bibnamefont {Knill}}, \bibinfo {author}
  {\bibfnamefont {R.}~\bibnamefont {Laflamme}}, \bibinfo {author}
  {\bibfnamefont {T.~F.}\ \bibnamefont {Havel}},\ and\ \bibinfo {author}
  {\bibfnamefont {D.~G.}\ \bibnamefont {Cory}},\ }\bibfield  {title} {\bibinfo
  {title} {Quantum simulation of a three-body-interaction hamiltonian on an nmr
  quantum computer},\ }\href {https://doi.org/10.1103/PhysRevA.61.012302}
  {\bibfield  {journal} {\bibinfo  {journal} {Phys. Rev. A}\ }\textbf {\bibinfo
  {volume} {61}},\ \bibinfo {pages} {012302} (\bibinfo {year}
  {1999})}\BibitemShut {NoStop}%
\bibitem [{\citenamefont {Nielsen}\ and\ \citenamefont
  {Chuang}(2011)}]{Nielsen:2011:QCQ:1972505}%
  \BibitemOpen
  \bibfield  {author} {\bibinfo {author} {\bibfnamefont {M.~A.}\ \bibnamefont
  {Nielsen}}\ and\ \bibinfo {author} {\bibfnamefont {I.~L.}\ \bibnamefont
  {Chuang}},\ }\href@noop {} {\emph {\bibinfo {title} {Quantum Computation and
  Quantum Information: 10th Anniversary Edition}}},\ \bibinfo {edition} {10th}\
  ed.\ (\bibinfo  {publisher} {Cambridge University Press},\ \bibinfo {address}
  {New York, NY, USA},\ \bibinfo {year} {2011})\BibitemShut {NoStop}%
\bibitem [{\citenamefont {Cerf}\ \emph {et~al.}()\citenamefont {Cerf},
  \citenamefont {Grover},\ and\ \citenamefont {Williams}}]{Cerf1998}%
  \BibitemOpen
  \bibfield  {author} {\bibinfo {author} {\bibfnamefont {N.~J.}\ \bibnamefont
  {Cerf}}, \bibinfo {author} {\bibfnamefont {L.~K.}\ \bibnamefont {Grover}},\
  and\ \bibinfo {author} {\bibfnamefont {C.~P.}\ \bibnamefont {Williams}},\
  }\bibfield  {title} {\bibinfo {title} {Nested quantum search and np-complete
  problems}\ }\href {https://doi.org/10.1103/PhysRevA.61.032303}
  {10.1103/PhysRevA.61.032303},\ \Eprint
  {https://arxiv.org/abs/http://arxiv.org/abs/quant-ph/9806078v1}
  {http://arxiv.org/abs/quant-ph/9806078v1} \BibitemShut {NoStop}%
\bibitem [{\citenamefont {Grover}(2005)}]{Grover2005}%
  \BibitemOpen
  \bibfield  {author} {\bibinfo {author} {\bibfnamefont {L.~K.}\ \bibnamefont
  {Grover}},\ }\bibfield  {title} {\bibinfo {title} {Fixed-point quantum
  search},\ }\bibfield  {journal} {\bibinfo  {journal} {Physical Review
  Letters}\ }\textbf {\bibinfo {volume} {95}},\ \href
  {https://doi.org/10.1103/PhysRevLett.95.150501}
  {10.1103/PhysRevLett.95.150501} (\bibinfo {year} {2005})\BibitemShut
  {NoStop}%
\bibitem [{\citenamefont {Yoder}\ \emph {et~al.}(2014)\citenamefont {Yoder},
  \citenamefont {Low},\ and\ \citenamefont {Chuang}}]{Yoder2014}%
  \BibitemOpen
  \bibfield  {author} {\bibinfo {author} {\bibfnamefont {T.~J.}\ \bibnamefont
  {Yoder}}, \bibinfo {author} {\bibfnamefont {G.~H.}\ \bibnamefont {Low}},\
  and\ \bibinfo {author} {\bibfnamefont {I.~L.}\ \bibnamefont {Chuang}},\
  }\bibfield  {title} {\bibinfo {title} {Fixed-point quantum search with an
  optimal number of queries},\ }\bibfield  {journal} {\bibinfo  {journal}
  {Physical Review Letters}\ }\textbf {\bibinfo {volume} {113}},\ \href
  {https://doi.org/10.1103/PhysRevLett.113.210501}
  {10.1103/PhysRevLett.113.210501} (\bibinfo {year} {2014})\BibitemShut
  {NoStop}%
\bibitem [{\citenamefont {Berry}\ \emph {et~al.}(2014)\citenamefont {Berry},
  \citenamefont {Childs}, \citenamefont {Cleve}, \citenamefont {Kothari},\ and\
  \citenamefont {Somma}}]{Berry2014}%
  \BibitemOpen
  \bibfield  {author} {\bibinfo {author} {\bibfnamefont {D.~W.}\ \bibnamefont
  {Berry}}, \bibinfo {author} {\bibfnamefont {A.~M.}\ \bibnamefont {Childs}},
  \bibinfo {author} {\bibfnamefont {R.}~\bibnamefont {Cleve}}, \bibinfo
  {author} {\bibfnamefont {R.}~\bibnamefont {Kothari}},\ and\ \bibinfo {author}
  {\bibfnamefont {R.~D.}\ \bibnamefont {Somma}},\ }\bibfield  {title} {\bibinfo
  {title} {Exponential improvement in precision for simulating sparse
  hamiltonians},\ }in\ \href {https://doi.org/10.1145/2591796.2591854} {\emph
  {\bibinfo {booktitle} {Proceedings of the 46th Annual {ACM} Symposium on
  Theory of Computing - {STOC} {\textquotesingle}14}}}\ (\bibinfo  {publisher}
  {{ACM} Press},\ \bibinfo {year} {2014})\BibitemShut {NoStop}%
\end{thebibliography}%
	
	\appendix
	
	\clearpage\widetext
	\section{Approximation of $n^*$ and $p_\eta$}
	\label{ref:A}
	Success probability $p(n,\epsilon)$ after $n$ shots measurement is 
	\begin{equation}\label{prob}
	p(n,\epsilon)=\langle\psi_0|\sin^{2}(\epsilon\Sigma)\cos^{2n-2}(\epsilon\Sigma)|\psi_0\rangle
	\end{equation}
	Output state after $n$ shots measurement is
	\begin{equation}
	|\psi(n,\epsilon)\rangle = \frac{\sin(\epsilon\Sigma)\cos^{n-1}(\epsilon\Sigma)|\psi_0\rangle}{\sqrt{\langle\psi_0|\sin^{2}(\epsilon\Sigma)\cos^{2n-2}(\epsilon\Sigma)|\psi_0\rangle}}
	\end{equation}
	The fidelity between output state and target state is
	\begin{align}\label{fidelity}
	f(n,\epsilon)= |\langle\psi(n,\epsilon)|\psi\rangle|^2 = 1-\frac{\langle\Sigma^6\rangle\langle\Sigma^2\rangle-\langle\Sigma^4\rangle^2}{\langle\Sigma^2\rangle^2}\left(\frac{n}{2}-\frac{1}{3}\right)^2\epsilon^4 + O(\epsilon^6)
	\end{align}
	where $\langle\Sigma^n\rangle = \langle\psi_0|\Sigma^n|\psi_0\rangle$ and
	\begin{equation}
	\langle\Sigma^6\rangle\langle\Sigma^2\rangle\ge\langle\Sigma^4\rangle^2
	\end{equation}
	The inequality above is Cauchy inequality and $\Sigma^2 = \Sigma$ saturates the inequality.
	The average measurement times $\bar n(\epsilon)$ to realize nonunitary operation is
	\begin{equation}\label{measure}
	\bar n(\epsilon)=\sum_{n\ge 1}p(n)n=\langle\psi_0|\sin^{-2}(\epsilon\Sigma)|\psi_0\rangle
	\end{equation}
	Eq.~(\ref{fidelity}) can be expanded at $\epsilon = 0$ to higher orders
	\begin{align}
		f(n,\epsilon) &=|\psi(n,\epsilon)|\psi|^2 = \frac{\langle\psi_0|\Sigma\sin(\epsilon\Sigma)\cos^{n-1}(\epsilon\Sigma)|\psi_0\rangle^2}{\langle\psi_0|\sin^{2}(\epsilon\Sigma)\cos^{2n-2}(\epsilon\Sigma)|\psi_0\rangle\langle\psi_0|\Sigma^{2}|\psi_0\rangle}\notag   = \frac{A\epsilon^2+B\epsilon^4+C\epsilon^6+D\epsilon^8}{A\epsilon^2+B\epsilon^4+C'\epsilon^6+D'\epsilon^8}
	\end{align}
	where
	\begin{align}
		A &= \langle\psi_0|\Sigma^2|\psi_0\rangle^2, \notag \\
		B &= \left(n-\frac{2}{3}\right)\langle\psi_0|\Sigma^4|\psi_0\rangle\langle\psi_0|\Sigma^2|\psi_0\rangle, \notag \\
		C &= \frac{1}{180}\left(45n^2-90n+48\right)\langle\psi_0|\Sigma^6|\psi_0\rangle\langle\psi_0|\Sigma^2|\psi_0\rangle+\left(\frac{n}{2}-\frac13\right)^2\langle\psi_0|\Sigma^4|\psi_0\rangle^2 \notag \\
		C' &= \frac{1}{90}(45n^2-60n+34)\langle\psi_0|\Sigma^6|\psi_0\rangle\langle\psi_0|\Sigma^2|\psi_0\rangle \notag \\  
		D &= \frac{1}{2520}\left(-105n^3+420n^2-588n+272\right)\langle\psi_0|\Sigma^8|\psi_0\rangle\langle\psi_0|\Sigma^2|\psi_0\rangle \notag \\ &+\frac{1}{360}\left(-45n^3+120n^2-108n+32\right)\langle\psi_0|\Sigma^6|\psi_0\rangle\langle\psi_0|\Sigma^4|\psi_0\rangle \notag \\
		D' &= \frac{1}{630}\left(-105n^3+315n^2-336n+124\right)\langle\psi_0|\Sigma^8|\psi_0\rangle\langle\psi_0|\Sigma^2|\psi_0\rangle
	\end{align}
	$n^*$ can be approximated by solving equation
	\begin{equation}\label{be}
		\frac{\langle\Sigma^6\rangle\langle\Sigma^2\rangle-\langle\Sigma^4\rangle^2}{\langle\Sigma^2\rangle^2}\left(\frac{n}{2}-\frac{1}{3}\right)^2\epsilon^4+\frac{\langle\Sigma^8\rangle\langle\Sigma^2\rangle-\langle\Sigma^6\rangle\langle\Sigma^4\rangle}{\langle\Sigma^2\rangle^2}\frac{(3n-2)(15n^2-30n+16)}{360}\epsilon^6 =2\eta
	\end{equation}
	where we omit $O(\eta^2)$. Next, we only keep terms like $(n\epsilon^2)^m$ in Eq.(\ref{be}) 
	\begin{equation}
	\frac{\langle\Sigma^6\rangle\langle\Sigma^2\rangle-\langle\Sigma^4\rangle^2}{\langle\Sigma^2\rangle^2}\frac{n^2\epsilon^4}{4} + \frac{\langle\Sigma^8\rangle\langle\Sigma^2\rangle-\langle\Sigma^6\rangle\langle\Sigma^4\rangle}{\langle\Sigma^2\rangle^2}\frac{n^3\epsilon^6}{8} = 2\eta
	\end{equation}
	Denote $x = n\epsilon^2$ and ignore cubic term
	\begin{equation}
		x = \frac{8\eta\langle\Sigma^2\rangle^2}{\langle\Sigma^6\rangle\langle\Sigma^2\rangle-\langle\Sigma^4\rangle^2}
	\end{equation}
	Hence we have
	\begin{equation}\label{n3}
		n^* = \sqrt{\frac{8\eta\langle\Sigma^2\rangle^2}{\langle\Sigma^6\rangle\langle\Sigma^2\rangle-\langle\Sigma^4\rangle^2}}\frac{1}{\epsilon^2}
	\end{equation}

	\section{Transfer matrix of coefficients}
	\label{sec:tm}
	 If there are $n$ ancillary qubits, transfer matrix $T_{2^n\times 2^n}$ can be written as
	 \begin{equation}
	 \left(
	 	\begin{pmatrix*}[r]
	 	2  & \dots & 2 \\
	 	\vdots & \ddots & \vdots \\
	 	2  & \dots & 2
	 	\end{pmatrix*}
	 	\begin{pmatrix*}[r]
	 	\langle\prod_{i=1}^n\cos^2(\epsilon\Sigma_i)\rangle & & \\
	 	& \ddots & \\
	 	& & \langle\prod_{i=1}^n\sin^2(\epsilon\Sigma_i)\rangle
	 	\end{pmatrix*}
	 	-I 
	 	\right)
	 	\begin{pmatrix*}[r]
	 	1 & & \\
	 	& \ddots & \\
	 	& & -1
	 	\end{pmatrix*}
	 \end{equation}
	 The characteristic polynomial is 
	 \begin{equation}
	 	|T-\lambda I| = (\lambda+1)^{2^n-2}\left(\lambda^2+(4s-2)\lambda+1\right)
	 \end{equation}
	 where $s = \langle\prod_{i=1}^n\sin^2(\epsilon\Sigma_i)\rangle$. Thus the eigenvalues of $T$ are
	 \begin{equation}
	 	-1,\quad -\left(\sqrt{s}-\sqrt{s-1}\right)^2,\quad -\left(\sqrt{s}+\sqrt{s-1}\right)^2.
	 \end{equation}
	 The corresponding $2^n-2$ eigenvectors of $-1$ are
	 \begin{align}
	 	\bm v_i = \left(-\frac{d_{i+1}}{d_1},0,\dots,1_{i+1},0\dots\right)^T,\quad i=1,2,\dots,2^n-2.
	 \end{align} 
	 And
	 \begin{align}
	 	\bm v_{-} &= \left(-\sqrt{\frac{s}{s-1}},-\sqrt{\frac{s}{s-1}},\dots,-\sqrt{\frac{s}{s-1}},1\right)^T \\
	 	\bm v_{+} &=\left(+\sqrt{\frac{s}{s-1}},+\sqrt{\frac{s}{s-1}},\dots,+\sqrt{\frac{s}{s-1}},1\right)^T
	 \end{align}
	 are eigenvectors of $-\left(\sqrt{s}-\sqrt{s-1}\right)^2$ and $-\left(\sqrt{s}+\sqrt{s-1}\right)^2$ respectively. Thus we have
	 \begin{equation}
	 	TX = X\mathrm{diag}\left(-1,-1,\dots,-1,-(\sqrt{s}-\sqrt{s-1})^2,-(\sqrt{s}+\sqrt{s-1})^2\right)
	 \end{equation}
	 where
	 \begin{equation}
	 	X = \begin{pmatrix}
	 	-\frac{d_2}{d_1} & -\frac{d_3}{d_1} & \dots & -\sqrt{\frac{s}{s-1}} & \sqrt{\frac{s}{s-1}} \\
	 	1 & 0 &\dots & -\sqrt{\frac{s}{s-1}} & \sqrt{\frac{s}{s-1}} \\
	 	\vdots & \vdots & \vdots & -\sqrt{\frac{s}{s-1}}  &\sqrt{\frac{s}{s-1}} \\
	 	0 & 0 & \dots & 1 & 1
	 	\end{pmatrix}
	 \end{equation}
   	$X^{-1}$ is a little hard to calculate but we will see later that only the last two rows of $X^{-1}$ contribute to the $\alpha^{(k)}_{2^n}$ which can be written as
   	\begin{equation}
   		\alpha^{(k)}_{2^n} = (0,0,\dots,0,1)T^k(1,1,\dots,1)^T
   	\end{equation}
	Substitute the eigenvalues and eigenvectors of $T$ into it and obtain
	\begin{align}
		\alpha^{(k)}_{2^n} = (-1)^k(0,0,\dots,0,\left(\sqrt{s}-\sqrt{s-1}\right)^{2k},\left(\sqrt{s}+\sqrt{s-1}\right)^{2k})X^{-1}(1,1,\dots,1)^T
	\end{align}
	The last two rows of $X^{-1}$ are
	\begin{equation}
		\begin{pmatrix*}[r]
		\frac{d_1}{2\sqrt{s(s-1)}} & \frac{d_2}{2\sqrt{s(s-1)}} & \dots & \frac{d_{2^n-1}}{2\sqrt{s(s-1)}} & \frac{1}{2} \\
		-\frac{d_1}{2\sqrt{s(s-1)}} & -\frac{d_2}{2\sqrt{s(s-1)}} & \dots & -\frac{d_{2^n-1}}{2\sqrt{s(s-1)}} & \frac{1}{2}
		\end{pmatrix*}
	\end{equation}
	Hence
	\begin{align}
		\alpha^{(k)}_{2^n} &= (-1)^k\sum_{i=1}^{2^n-1}\left(\left(\sqrt{s}-\sqrt{s-1}\right)^{2k}\frac{d_i}{2\sqrt{s(s-1)}}-\left(\sqrt{s}+\sqrt{s+1}\right)^{2k}\frac{d_i}{2\sqrt{s(s-1)}}\right) \notag \\
		&+(-1)^k\frac{1}{2}\left(\left(\sqrt{s}+\sqrt{s-1}\right)^{2k}+\left(\sqrt{s}-\sqrt{s-1}\right)^{2k}\right) \notag \\
		&=\frac{1}{2\sqrt{s}}\left(\left(\sqrt{s}+\sqrt{s-1}\right)^{2k+1}+\left(\sqrt{s}-\sqrt{s-1}\right)^{2k+1}\right)
	\end{align}
	Thsu the success probability $p_1(s,k) = s\alpha^{(k)}_{2^n}$ is
	\begin{equation}
	p_1(s,k) = \frac{1}{4}\left[\left(\sqrt{s}+\sqrt{s-1}\right)^{2k+1}+\left(\sqrt{s}-\sqrt{s-1}\right)^{2k+1}\right]^2
	\end{equation}
	
	\section{$R_i$}
	\label{sec:ri}
	
		\begin{align}\label{reflection}
	R_i &= U_iG_{i-1}H_{i+1}\left[2|0\rangle\langle 0|_{\rm{anc}}\otimes|\psi_{i-1}\rangle\langle\psi_{i-1}\otimes|0\rangle\langle 0|_{{i+1}}-I\right]H_{i+1}G_{i-1}^\dagger U_i^\dagger \notag \\
	&= U_iG_{i-1}H_{i+1}\left\{|0\rangle\langle 0|_{\rm{anc}}\otimes\left[(2|\psi_{i-1}\rangle\langle\psi_{i-1}|-I)\otimes|0\rangle\langle 0|_{i+1} - |1\rangle\langle 1|_{i+1}\right] -|1\rangle\langle 1|_{\rm{anc}}\otimes I\right\}H_{i+1}G_{i-1}^\dagger U_i^\dagger \notag \\
	&= U_iH_{i+1}\left\{|0\rangle\langle 0|_{\rm{anc}}\otimes G_{i-1}R_{i-1}G_{i-1}^\dagger\otimes |0\rangle\langle 0|_{i+1} - |0\rangle\langle 0|_{\rm{anc}}\otimes I\otimes |1\rangle\langle 1|_{i+1} - |1\rangle\langle 1|_{\rm{anc}}\otimes I\right\}H_{i+1}U_i^\dagger
	\end{align}
	where $\epsilon_i$ in $U(\Sigma_i,\epsilon_i)$ is defined by
	\begin{equation}\label{tstar}
	t^*_i =
	\frac{\mathrm{Tr}\left[\cos^2(\epsilon_i\Sigma_i)\prod_{j=1}^{i-1}\sin^2(\epsilon_j\Sigma_j)\right]}{2\mathrm{Tr}\left[\prod_j^{i-1}\sin^{2}(\epsilon_j\Sigma_j)\right]}
	\end{equation}
	Substitute Eq.(\ref{sigma}) into Eq.(\ref{tstar}) and we have
	\begin{equation}
	t^*_i = \frac{1}{2}\left[1+\cos\left(\epsilon_i+\epsilon_ie^{-2J_i\tau})\right)\cos\left(\epsilon_i-\epsilon_ie^{-2J_i\tau})\right)\right]
	\end{equation}
			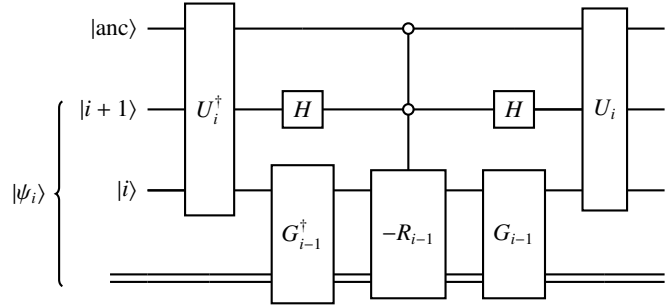
\begin{figure}[h]
		\centering
		\begin{adjustbox}{width = 0.5\textwidth}
			\begin{quantikz}
				&[0.01cm] &\lstick{\ket{\rm{anc}}} & \gate[wires = 3]{U_i^\dagger} &  \qw & \octrl{1}  & \qw & \gate[wires = 3]{U_i} & \qw \\
				\lstick[wires=3]{\ket{\psi_{i}}}&[0.2cm]	&\lstick{\ket{i+1}} &                   &\gate{H}                 & \octrl{1} & \gate{H} & \qw & \qw  \\
				&[0.01cm]	&\lstick{\ket{i}} & \qw & \gate[2,cwires={2}]{G^\dagger_{i-1}} &\gate[2,cwires={2}]{-R_{i-1}} & \gate[2,cwires={2}]{G_{i-1}} & &\qw \\
				&[0.01cm] &\cw	&\cw & & &  & \cw &\cw 
			\end{quantikz}
		\end{adjustbox}
		\caption{The reflection operator $R_i$ can be constructed recursively. }\label{reflectioncircuit}
	\end{figure}

\end{document}